\documentclass[%
 reprint,
 superscriptaddress,
 amsmath,amssymb,
 aps,
 prx,
]{revtex4-2}

\usepackage{graphicx}
\usepackage{dcolumn}
\usepackage{bm}
\usepackage{hyperref}

\usepackage{float}
\usepackage{array}

\usepackage{listings}
\setcitestyle{numbers,square}

\begin{document}


\title{Towards Verifiable and Self-Correcting AI Physicists \\for Quantum Many-Body Simulations}

\author{Ken Deng}
\affiliation{Department of Physics, Tsinghua University, Beijing 100084, China}

\author{Xiangfei Wang}
\affiliation{Department of Physics, Tsinghua University, Beijing 100084, China}

\author{Guijing Duan}
\affiliation{Department of Physics and Beijing Key Laboratory of Opto-electronic Functional Materials \& Micro-nano Devices, Renmin University of China, Beijing 100872, China}

\author{Chen Mo}
\affiliation{School of Physics, Sun Yat-sen University, Guangzhou 510275, China}

\author{Junkun Huang}
\affiliation{School of Gifted Young, University of Science and Technology of China, Hefei 230026, China}

\author{Runqing Zhang}
\affiliation{China Mobile (Suzhou) Software Technology Co., Ltd., Suzhou 215163, China}

\author{Ling Qian}
\affiliation{China Mobile (Suzhou) Software Technology Co., Ltd., Suzhou 215163, China}

\author{Zhiguo Huang}
\affiliation{China Mobile (Suzhou) Software Technology Co., Ltd., Suzhou 215163, China}

\author{Jize Han}
\affiliation{China Mobile (Suzhou) Software Technology Co., Ltd., Suzhou 215163, China}

\author{Di Luo}
\email{diluo1000@gmail.com}
\affiliation{Department of Physics, Tsinghua University, Beijing 100084, China}
\affiliation{Institute of Advanced Study, Tsinghua University, Beijing 100084, China}

\date{\today}

\begin{abstract}
While large language models (LLMs) promise to revolutionize automated scientific discovery, their application in rigorous real-world physical research is stalled by two critical barriers: a lack of realistic evaluation benchmarks and systemic LLM hallucinations. Here, we address both problems. We introduce QMP-Bench, a pioneering end-to-end research-level benchmark in quantum many-body simulation consisting of $100$ tasks extracted from $21$ high-impact prestigious journals, presenting a challenge even for current frontier LLMs. To establish a paradigm for reliable and transparent AI physicists, we present PhysVEC, a multi-agent framework that enforces self-verifiable and error correction in AI research. PhysVEC seamlessly integrates programming and scientific verifiers to guarantee coding correctness and principle-based physical validity, yielding interpretable evidence and error correction at each step. PhysVEC significantly outperforms existing LLM baselines on various scenarios in QMP-Bench and presents a favorable inference-time scaling, successfully transforming unreliable AI generations into accurate physical reproductions, paving a robust and trustworthy path towards future automated scientific discovery.
\end{abstract}

\maketitle


\section{\label{sec1}Introduction}

Automated scientific discovery has emerged as a transformative force across various frontier research fields, from accelerating machine learning \cite{TheAIScientist,AIresearcher,DeepCode}, material synthesis \cite{AutonomousMaterialLab,AIatomicmicroLabs}, to reshaping experimental physics \cite{AINanoLabs,AIquancompLabs,AgentSensor} and theoretical physics \cite{DeepInflation,LLMFeynman,AgentQcode,QcircuitBench,ElAgente,AutoQuanSim,PhysMaster}. This surge is fueled by both (1) the improvement in capacity of large language models \cite{CoTprompt,ToT,r1,RAGLewis,SelfRAG} and (2) the advancement of the agent paradigm \cite{ReAct,PAL,SelfRefine,SWE-agent,MCP}. Such systems dramatically lower the barrier for researchers to adopt diverse tools and compress the iteration cycle between theoretical insight and empirical verification. Beyond efficiency, automated systems open new possibilities for cross-disciplinary discovery, leveraging the LLMs' capability to connect and integrate insights across disciplines that human researchers might rarely bridge \cite{SciencePedia}. Specifically, AI-driven systems find a particularly natural application in scientific simulations. Several recent studies have demonstrated strong performance of LLM-based agents in this field \cite{QcircuitBench,ElAgente,AutoQuanSim,PhysMaster}.

Despite this progress, critical gaps remain before AI-driven systems can be reliably deployed in practical research. In scientific simulations, ensuring both programming correctness and physical validity is essential. Although LLM-based agents have shown impressive performance in curated problems and exam-level tasks \cite{SciBench,GPQA,PhysSupernova,ElAgente,AutoQuanSim,CriPt,SciCode}, reproducing the results of peer-reviewed journal articles presents a qualitatively more difficult challenge. Moreover, LLM-generated scientific scripts often suffer from hallucinations, including domain-specific errors in both code syntax and physical configurations. Both challenges above make verification and error correction indispensable in agent systems. Existing approaches to this problem fall into two types, each with fundamental limitations. Expert-curated gold answers ensure correctness, but require substantial human effort and are often impractical for realistic research tasks where the definitive ground truth cannot be specified in advance \cite{TPBench,CriPt}. LLM-as-a-judge approaches enable scalable automated evaluation but inherit the very hallucinations they are meant to detect, while producing judgments that are difficult for human researchers to verify \cite{AIresearcher,DeepScientist,JudgeLLMjudge}. 

In this work, we first present QMP-Bench, to our knowledge the first end-to-end research-level benchmark dataset focusing on \textbf{Q}uantum \textbf{M}any-Body \textbf{P}hysics. The dataset comprises $100$ tasks drawn from $21$ high-impact published articles. Unlike evaluations based on human-curated intermediate problems \cite{TPBench,CriPt,SciCode}, QMP-Bench targets actual end-to-end scientific simulation tasks. Quantum many-body simulations, which model interacting quantum systems exhibiting complex emergent phenomena, are central to frontier physics research and are technically demanding, making them a suitable testbed for AI physicists. We demonstrate that QMP-Bench poses a formidable challenge to current frontier LLMs (\texttt{GPT-5.1}, \texttt{Gemini 2.5 Flash}, \texttt{Qwen3-Max}, and \texttt{Claude Sonnet 4}), which struggle with these tasks and exhibit notably poor performance. To address this critical limitation, we introduce PhysVEC, a multi-agent AI physicist framework for automated self-\textbf{V}erification and \textbf{E}rror \textbf{C}orrection, equipped with structural tools for domain-specific scientific computations. By evaluating QMP-Bench, we show that PhysVEC consistently outperforms baseline approaches across all models and topics. Furthermore, PhysVEC exhibits clear inference-time scaling effects, achieves rigorous programming and physical correctness, and ultimately produces faithful and interpretable research results.

\section{QMP-Bench Dataset and PhysVEC framework}

\subsection{Challenges for LLM Agents in Scientific Research}

Despite the remarkable progress of LLMs in facilitating scientific reasoning and accelerating domain discoveries, deploying them into practical scientific simulations faces two critical bottlenecks. First, the community lacks realistic and domain-specific benchmark datasets required to evaluate continuous computational workflows. Second, existing frameworks broadly fail to establish mechanisms that operate reliably and provide transparent evidence, which are inherently required to faithfully output correct physical results. Resolving these intertwined challenges is critical for the future of AI-driven physical research.

To evaluate the capabilities of LLMs in scientific reasoning, the community has developed numerous benchmarks, yet a gap with real research persists. Early evaluation systems, such as SciBench \cite{SciBench}, GPQA \cite{GPQA}, PHYSICS \cite{PHYSICS}, and Physics Supernova \cite{PhysSupernova}, have provided excellent testbeds for assessing at the exam-level or Olympiad-level. Recognizing the necessity for more advanced evaluations, subsequent studies like CriPt \cite{CriPt} and TPBench \cite{TPBench} have introduced benchmarks that are closer to professional physics research. However, these remain predominantly human-curated and simplified tasks, unable to mirror the long-horizon nature of research implementation. More recently, benchmarks such as PRBench \cite{PRBench} and PRL-Bench \cite{PRL-Bench} have proposed physical research tasks derived directly from published articles, sharing a highly similar objective with our approach. Nevertheless, evaluating the ability of AI-driven systems to execute complex computational physics simulations remains largely unaddressed. A critical gap exists for benchmarks that concentrate specifically on high-barrier domains (e.g. quantum many-body physics) and that rigorously demand the utilization of community-recognized third-party libraries (e.g. ITensors, NetKet) to accomplish tasks.

In standard human research workflows, systematic verification and rigorous error correction are core mechanisms for ensuring scientific interpretability and validity, which have not been explicitly discussed in recent AI scientist designs. In AI-driven research, previous works have introduced formal systems (e.g. Lean) into mathematical and physical theorem proving, making the logical steps strictly verifiable \cite{AlphaProof,Lean4Physics,AxProver}. However, integrating such formal languages into complex and numerical scientific simulations remains highly impractical today. Consequently, most existing agent systems for scientific simulation tasks default to ReAct-style iterative architectures \cite{ReAct}. These frameworks suffer from several critical limitations. For verification, LLM-generated scripts are often structurally diverse and highly variable across models and contexts, severely complicating the implementation of systematic and automated checking. As a result, the scientific outputs are usually opaque and difficult for human reviewers to scrutinize, leading to highly uncertain physical validity. Furthermore, existing conventional architectures exhibit pronounced weaknesses in error correction. Conventional iterative loops typically surface and return only the first runtime error encountered at a time, resulting in an exceptionally sluggish refinement trajectory. More crucially, for scientific validity errors, identifying and localizing the underlying cause remains a challenge.

\subsection{QMP-Bench: End-to-end Benchmark of High Impact Research Paper on Quantum Many-body Simulations}

\subsubsection{Dataset construction}

\begin{figure*}[htbp] 
    \centering
    \includegraphics[width=0.65\textwidth]{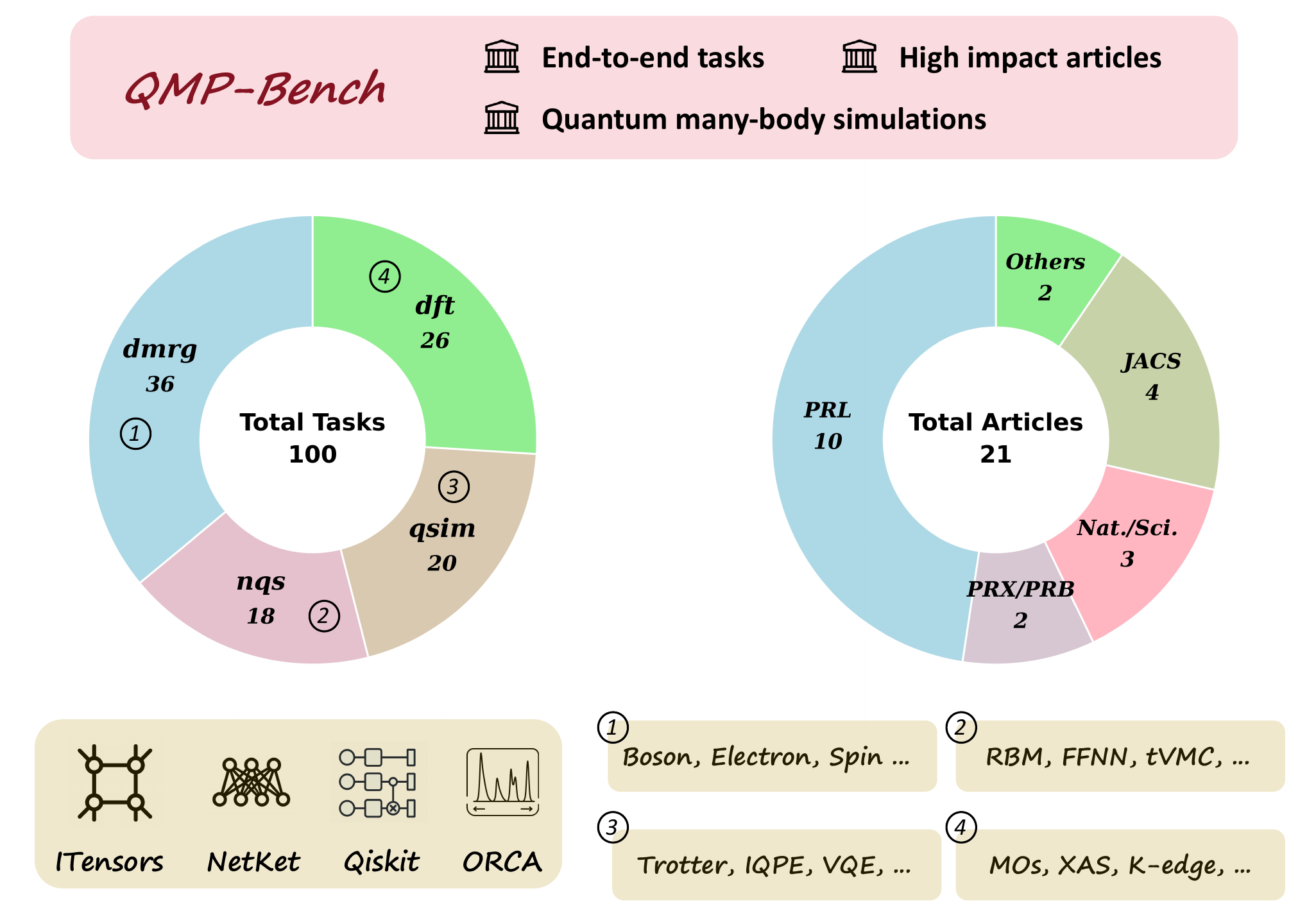}
    \caption{Overview of the QMP-Bench dataset. The benchmark is designed around three core characteristics: end-to-end tasks, curation from high-impact articles, and a focus on quantum many-body physics. QMP-Bench contains a total of $100$ computational tasks extracted from $21$ highly influential articles. The dataset broadly covers essential physical systems and advanced algorithms, spanning tensor networks, neural quantum states, quantum circuit simulations, and density-functional theory calculations. These benchmark tasks represent a comprehensive range of numerical studies that can be implemented using standard modern software packages, specifically ITensors, NetKet, Qiskit, and ORCA. }
    \label{fig:dataset}
\end{figure*}

We first introduce QMP-Bench, a dataset of $100$ research-level tasks for quantum many-body simulations. Unlike benchmarks built around human-curated intermediate artifacts, QMP-Bench provides end-to-end tasks that require reproducing results directly from the original article. Quantum many-body physics studies systems of interacting quantum particles whose collective behavior gives rise to emergent phenomena such as superconductivity and topological phases, and whose simulation is inherently challenging due to exponential complexity. Quantum many-body simulations are both scientifically important and practically challenging to carry out in research, making this domain a valuable testbed for AI-driven research. To our knowledge, QMP-Bench is the first high-impact and research-level benchmark to systematically evaluate end-to-end scientific research in quantum many-body physics.

Fig.~\ref{fig:dataset} demonstrates the composition of this dataset. QMP-Bench includes $36$ tensor‑network problems (denoted \texttt{dmrg}) with ITensors \cite{ITensors}, $18$ neural‑network ansatz problems (\texttt{nqs}) with NetKet \cite{NetKet,NetKet3}, $20$ quantum circuit problems (\texttt{qsim}) with Qiskit \cite{Qiskit}, and $26$ density-functional theory problems (\texttt{dft}) with ORCA \cite{ORCA} (see more details in Appendix~\ref{appendixsec-packages}). These tasks are drawn from $21$ published high impact articles. As summarized in the figure, QMP-Bench covers a range of canonical models and algorithms in quantum many-body simulations. The list of articles is given in Appendix~\ref{appendixsec-articlelist}. 

We construct the QMP-Bench dataset as follows. For each chosen article, we select figures that can be explicitly computed with ITensors, NetKet, Qiskit, or ORCA. In the QMP-Bench dataset, each task consists of two components: (1) The .tex source file of the corresponding article collected from arXiv. (2) A short user request, in which we explicitly specify the figure in the article to be reproduced. If a figure involves several parameter choices or extensive scans, the user request restricts the task to reproducing only a few representative cases to avoid redundant computation. The Author agent then starts to construct the simulation script based on the article’s .tex file and the corresponding user request.

\subsubsection{Baseline performances}

\begin{figure*}[htbp] 
    \centering
    \includegraphics[width=0.87\textwidth]{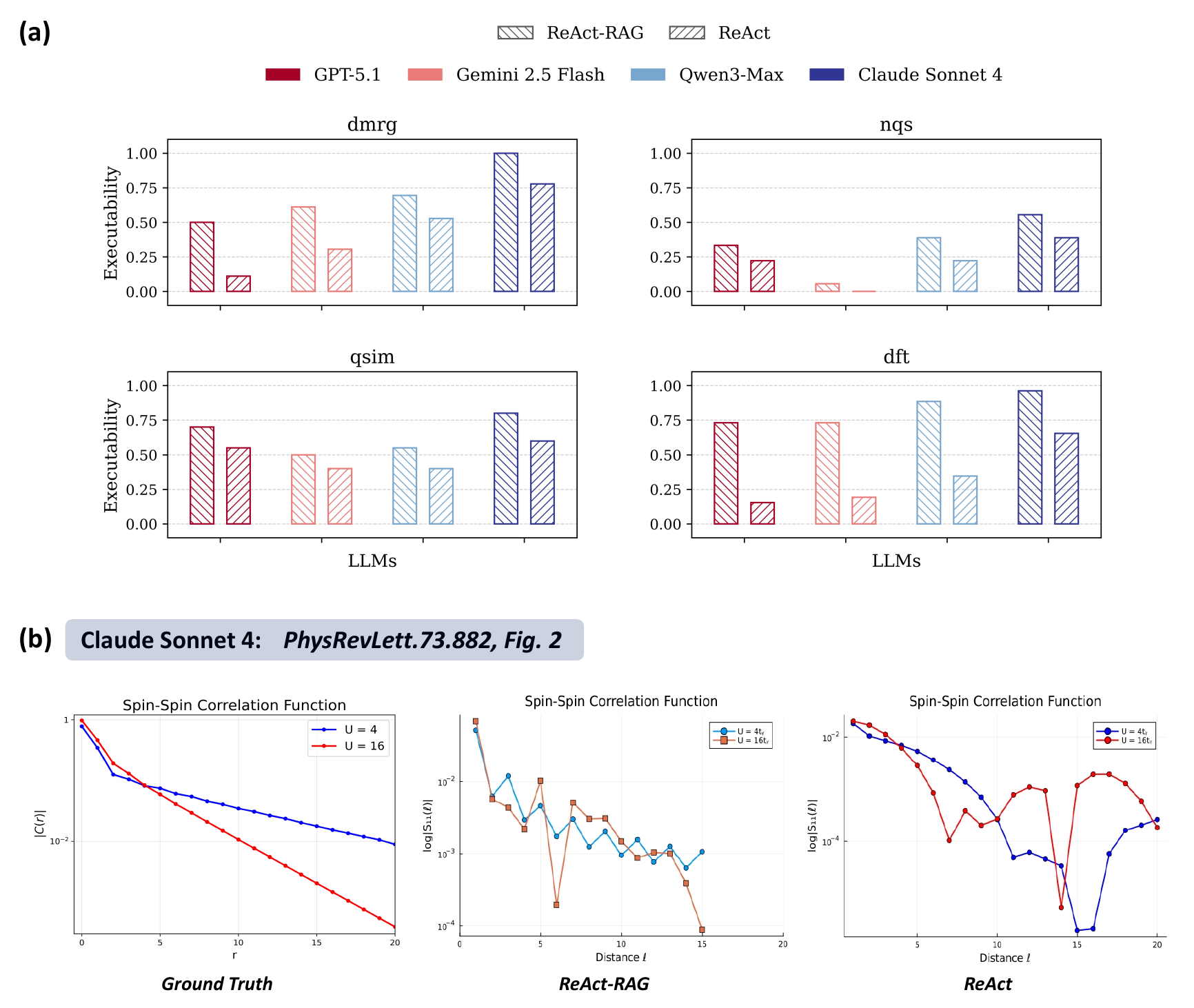}
    \caption{Limitations of frontier LLMs on QMP-Bench revealing significant challenges in automated quantum many-body simulations under conventional approaches (ReAct-RAG and ReAct). (a) Executability (pass@8 programming correctness) of the final generated scripts across four task categories. The results highlight the intrinsic difficulty of QMP-Bench. Across all tasks, the proportion of error-free scripts generated by standard ReAct architecture generally falls below $50\%$, and rarely exceeds $75\%$ even when utilizing ReAct-RAG. Notably, while \texttt{Claude Sonnet 4} achieves the overall best performance via ReAct-RAG, it still yields an executability of only about $50\%$ in \texttt{nqs} tasks. (b) Failure mode analysis of the highest-performing model (\texttt{Claude Sonnet 4}) on reproducing the spin-spin correlation function of a two-chain Hubbard model \cite{PhysRevLett.73.882}. Both baseline methods deviate critically from the ground truth. Specifically, ReAct-RAG configures an incorrect particle number for the half-filling condition, whereas the ReAct workflow inherently misdefines the hopping terms in the model's Hamiltonian.}
    \label{fig:baseline_result}
\end{figure*}

To establish the inherent difficulty of the QMP-Bench dataset, we benchmark the capabilities of frontier LLMs utilizing conventional ReAct-style architectures. We observe that these baselines exhibit profound deficiencies in both programming correctness and physical validity, struggling to produce script that is either executable or scientifically sound. In the baseline workflows, the agent ingests a raw research article to generate an initial computational script, which subsequently undergoes an iterative execution-repair loop based on the runtime feedback (denoted as ``ReAct''). Furthermore, to simulate a well-equipped computational researcher, we also consider the workflow empowering the LLM to retrieve guidance from a local knowledge base (comprising the official documentation and manuals of the required numerical packages, see more details in Appendix~\ref{appendixsec-reference}) during each repair iteration (denoted as ``ReAct-RAG'').

We first assess the programming correctness (``executability'') of the scripts from these baselines, defined as the proportion of scripts that run without runtime exceptions within a maximum of eight iterations (pass@8). The quantitative results across four distinct topics, presented in Fig.~\ref{fig:baseline_result} (a), starkly underscore the formidable challenge posed by QMP-Bench. Despite the iterative debugging mechanisms, scripts generated by the ReAct architecture predominantly fail to execute, with success rates generally below $50\%$. Even when utilizing external documentation retrieval in the ReAct-RAG setup, the executability rarely exceeds the $75\%$ threshold. Although \texttt{Claude Sonnet 4} emerges as the model that is the most capable overall under the ReAct-RAG framework, it still struggles to achieve an executability of only about $50\%$ in the \texttt{nqs} tasks. Driven by the extensive use of third-party numerical libraries and their inherent compatibility hurdles, this widespread inability to resolve syntactic or library-calling errors highlights a critical bottleneck in conventional coding agents.

\begin{figure*}[htbp] 
    \centering
    \includegraphics[width=0.75\textwidth]{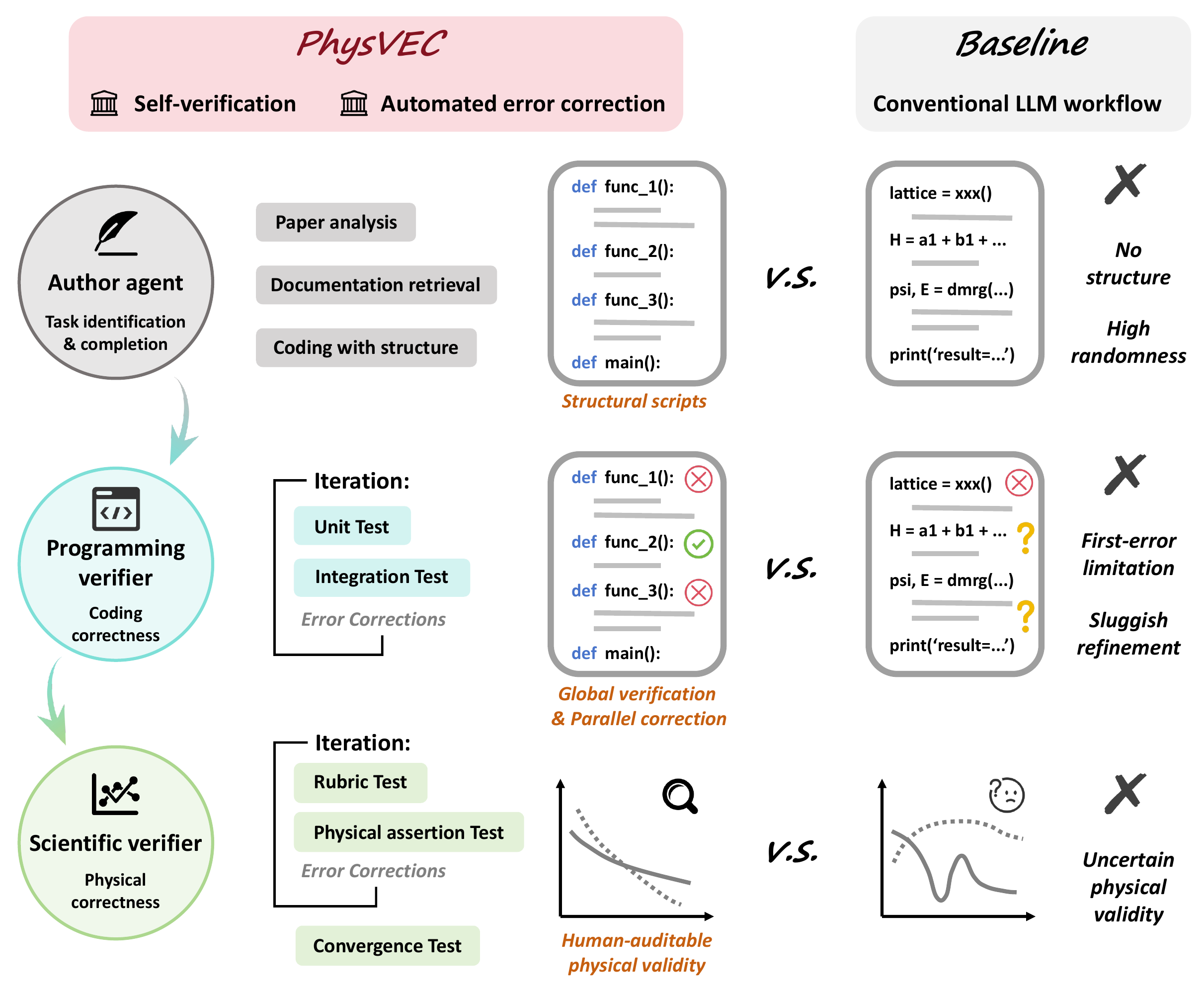}
    \caption{Overview of the PhysVEC framework. PhysVEC is a multi-agent AI physicist for quantum many-body simulations with self-verification and automated error correction design. The framework consists of three key components, with its core innovation lying in the two verifier agents. Specifically, the Programming verifier is responsible for checking and ensuring the programming correctness of the generated scripts through unit tests and integration tests. Subsequently, the Scientific verifier examines and ensures the physical validity of the scripts via physics-informed checks. Compared to conventional LLM baselines (e.g. simple execution-repair loops), the Author agent in PhysVEC ensures the generation of structurally stable scripts that are independent of specific LLMs. Based on this structural foundation, the two verifiers realize a more efficient refinement process, ultimately yielding human-auditable final results.}
    \label{fig:workflow}
\end{figure*}

Beyond programming failures, the baseline architectures also suffer from a severe lack of physical validity, resulting in entirely erroneous physics. Fig.~\ref{fig:baseline_result} (b) provides a failure mode demonstration of the highest-performing \texttt{Claude Sonnet 4} on a \texttt{dmrg} task. This task requires reproducing the spin-spin correlation function of a two-chain Hubbard model \cite{PhysRevLett.73.882}. Both baseline approaches critically deviate from the ground truth (Fig.~\ref{fig:baseline_result} (b) left). Although their scripts eventually run without throwing programming exceptions, they harbor physical misunderstandings. Specifically, the ReAct-RAG generated script incorrectly configures the total particle number, fundamentally violating the required half-filling condition of the target quantum system. On the other hand, the ReAct script misdefines the hopping terms within the model's Hamiltonian, which turns out to be unphysically non-Hermitian. These scientifically flawed outcomes clearly demonstrate that even when equipped with frontier LLMs, current baseline architectures exhibit a profound gap in ensuring faithful reproduction, fundamentally demanding a more rigorous and physics-aware paradigm.

Consequently, resolving the vulnerability of current architectures requires a tailored agent framework that goes beyond generic execution feedback. This new paradigm must explicitly incorporate robust self-verification and efficient error correction mechanisms into its core workflow. Only through such syntactically rigorous and physics-aware validation loops can LLMs reliably conquer the challenges encountered in QMP-Bench, paving the way for deployment in accelerating real-world scientific simulations.

\begin{figure*}[htbp] 
    \centering
    \includegraphics[width=0.75\textwidth]{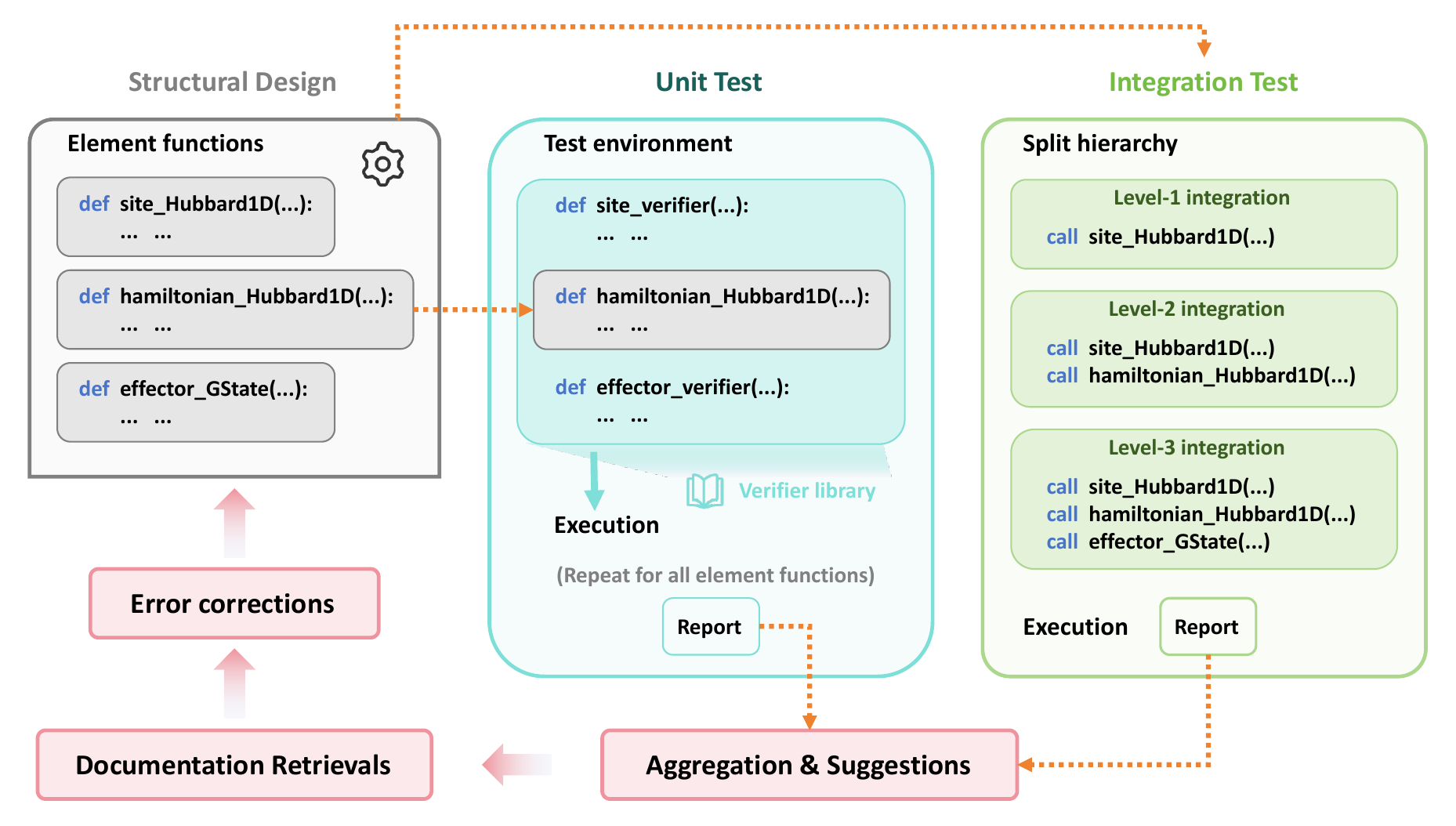}
    \caption{Verification and error correction via unit tests and integration tests. In the generated script, the Author agent defines a set of element functions that are subsequently called to perform the computations (gray block). Then the Programming verifier conducts unit tests (blue block) and integration tests (green block) for all element functions. The Programming verifier aggregates the reports and implements corrections accordingly (red blocks). }
    \label{fig:unit_integ}
\end{figure*}

\subsection{PhysVEC: A Verifiable and Error Correctable AI Physicist Framework}

Here, we introduce the PhysVEC framework. Equipped with programming tests and scientific tests, PhysVEC systematically verifies LLM-generated results and performs iterative error correction to achieve accurate and interpretable scientific results, as shown in the left column of Fig.~\ref{fig:workflow} (a). PhysVEC consists of three cooperating agents: an \emph{Author agent}, a \emph{Programming verifier}, and a \emph{Scientific verifier}. The Author agent analyzes the input paper and completes the task. The Programming verifier then conducts programming tests to verify programming correctness and fix syntax errors. Finally, the Scientific verifier performs scientific tests to verify physical validity via physics-informed checks and outputs the reproduced results (see more details about the PhysVEC pipeline in the Supplemental Material). 

The Author agent identifies and completes the target task from the original research article. It first analyzes the paper, conducts free-form planning, consults relevant library documentation or manuals, and then iteratively refines the script until it conforms to a predefined structure. Concretely, the script is organized into reusable blocks, denoted as ``element functions'' (e.g., constructing the lattice, defining the Hamiltonian, etc.), which can be systematically composed in subsequent computations, as illustrated in Fig.~\ref{fig:unit_integ} (gray block). Explicit modularization allows each block to be independently inspected and revised, which enables standardized and systematic verification and efficient error correction, laying the foundation for the subsequent procedures. 

The Programming verifier examines the syntax correctness of the script and performs corresponding corrections to guarantee its executability. The agent performs two types of test: unit tests and integration tests, which are usually missed in practical scientific simulations. In unit tests (Fig.~\ref{fig:unit_integ} blue block), the agent examines each element function definition in isolation. The agent constructs an environment around the function to be tested, and executes it to reveal any errors in programming within the target element function (see more details in Appendix~\ref{appendixsec-verifierlib}). In integration tests (Fig.~\ref{fig:unit_integ} green block), the agent first decomposes the computation process into multiple levels and then all levels are executed to identify any compatibility issues. The results of these levels are also used to double-check the judgments made in the unit tests. Subsequently, the Programming verifier aggregates all diagnostic reports and corrects the detected issues. This verification-and-correction procedure is run iteratively until the full code executes without errors. 

The Scientific verifier performs sanity checks on the script that has passed the programming test, and outputs the reproduced figure of the original article upon successful validation. It conducts three types of physics-informed verification: the rubric test \cite{PaperBench}, the physical assertion test, and the convergence test. In the rubric test, the agent inspects the generated script against a manually curated rubric, checking the physical system definitions and the numerical solver configurations. In the physical assertion test, the agent leverages physics principle-based assertions (see more details in Fig.~\ref{fig:scittest} (a) and Appendix~\ref{sec:append-scitest}) to adapt the original script into multiple test variants, and assesses physical validity against the expected outcomes of each assertion. It further performs error corrections accordingly based on the diagnostic evidence provided by these tests. Finally, in the convergence test, the agent progressively increases computation parameters (e.g., the number of sweeps in DMRG settings) until it determines that convergence is achieved, ensuring the robustness of the research results.

\begin{figure*}[htbp]
    \centering
    \includegraphics[width=0.95\linewidth]{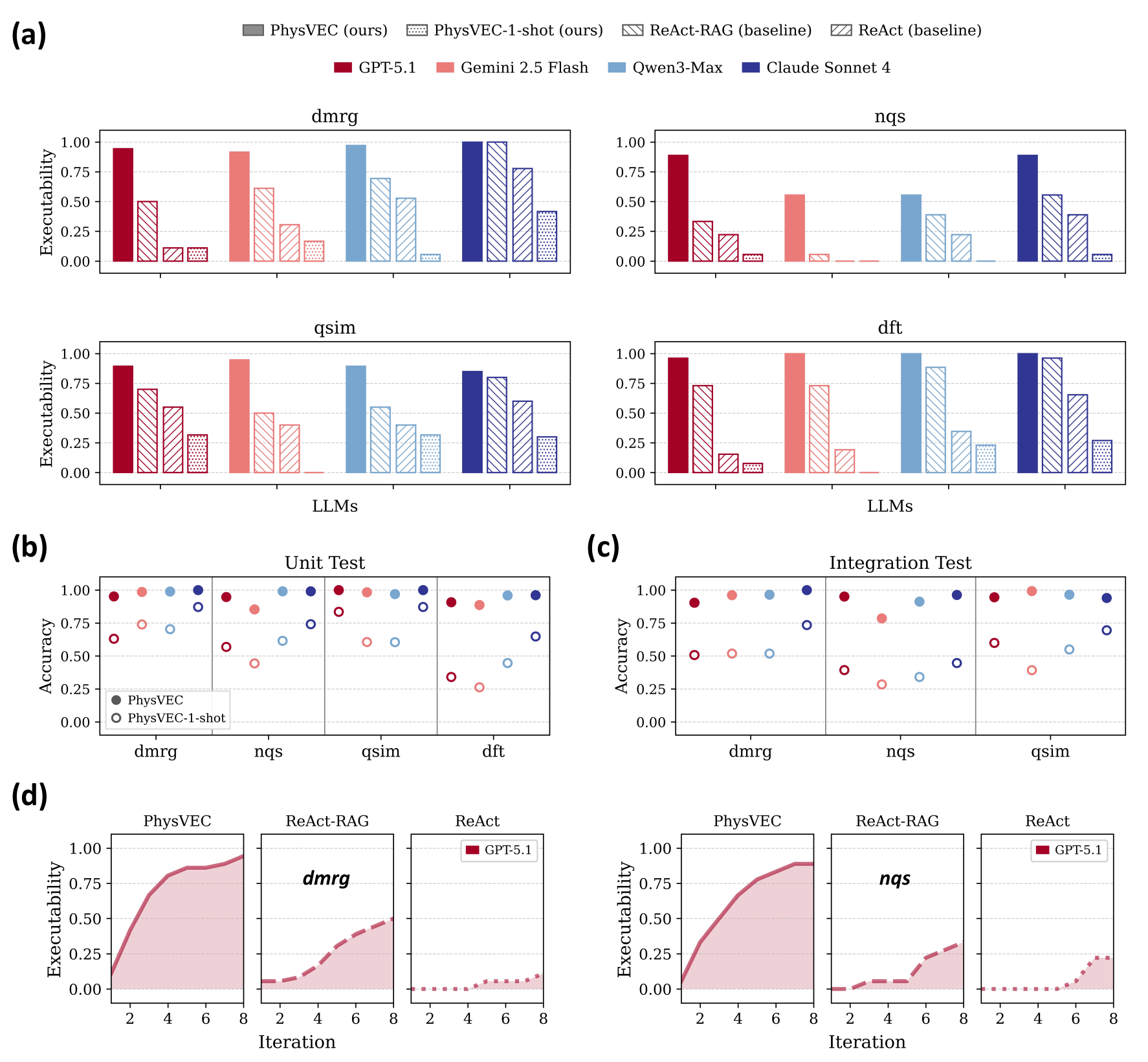}
    \caption{Evaluation of programming correctness (pass@8) demonstrating the substantial performance improvements achieved by PhysVEC compared to baseline frameworks. (a) Comparison of script executability across four frontier LLMs using our PhysVEC framework and three baselines (PhysVEC-1-shot, ReAct-RAG, and ReAct). The results explicitly show that PhysVEC brings significant enhancements to the executability of the generated scripts for all tested LLMs. This robust improvement is particularly notable in the \texttt{nqs} tasks, where baseline performances were observed to be the poorest. Furthermore, under the PhysVEC framework, the executability approaches nearly $100\%$ for several model-topic combinations. (b) (c) The accuracy measured in unit tests and integration tests before (hollow markers, PhysVEC-1-shot) and after (solid markers, PhysVEC) the iterative verification and error correction. These two panels provide a finer-grained perspective confirming the programming correctness improvement brought by PhysVEC. Note that integration tests are not applicable to the \texttt{dft} input files. (d) The proportion of correctly running scripts (executability) as a function of iterations. The curves display the trajectories of \texttt{GPT-5.1} on the \texttt{dmrg} and \texttt{nqs} topics across different frameworks, while other model-topic combinations exhibit similar behaviors. The dynamics reveal that the PhysVEC framework systematically converges to a much higher saturation value of executability with significantly fewer iteration steps.}
    \label{fig:programtest}
\end{figure*}

\section{Results}

This section details the comprehensive performance enhancements achieved by PhysVEC for quantum many-body simulations on QMP-Bench. We evaluate this framework from two perspectives: programming correctness and physical validity. In the PhysVEC framework, the Programming verifier resolves syntactical errors to make scripts executable, while the Scientific verifier performs physics-informed checks to guarantee physically faithful reproductions.

\subsection{Programming verification and error correction}

\begin{figure*}
    \centering
    \includegraphics[width=0.8\textwidth]{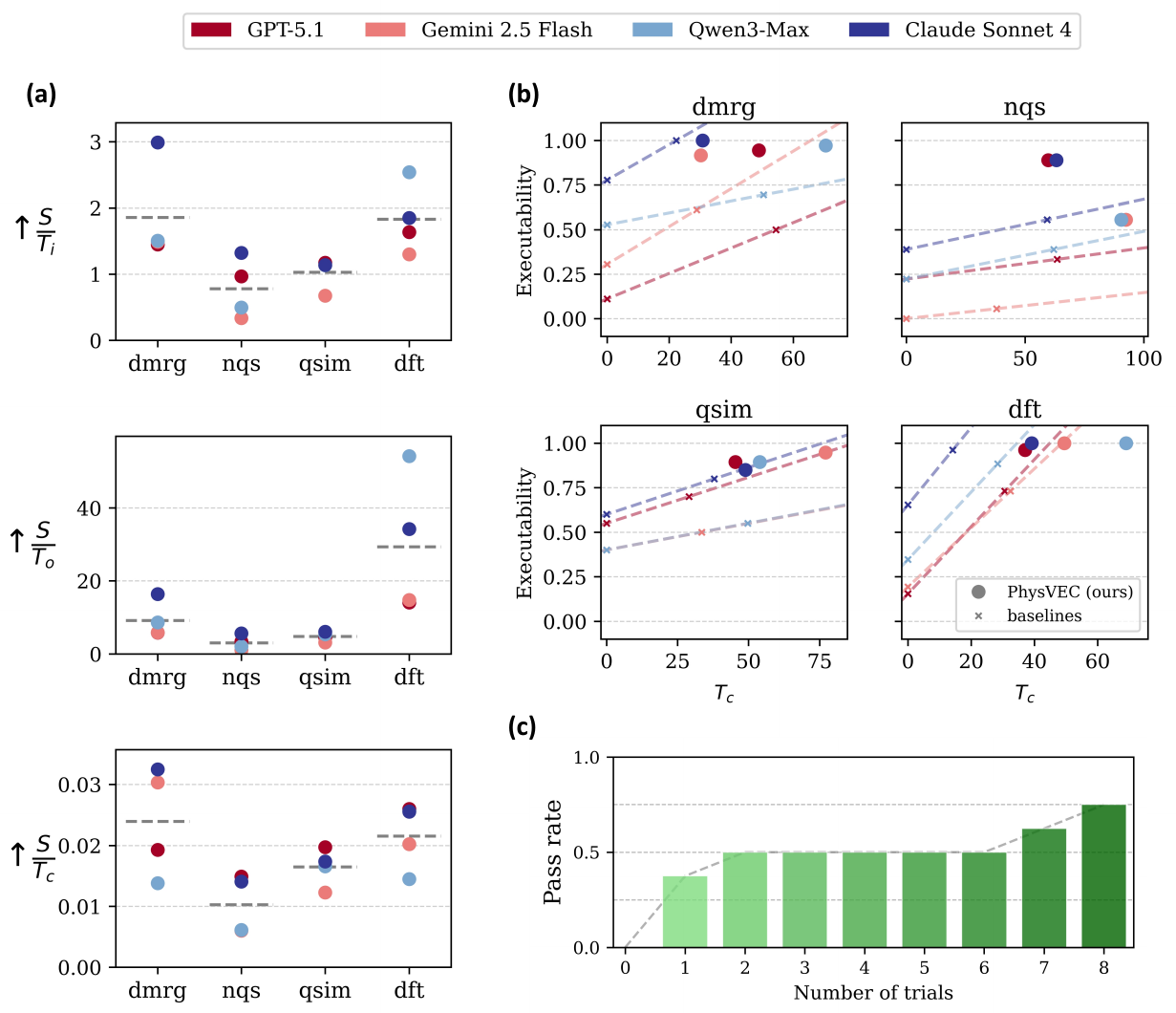}
    \caption{Detailed behavioral analysis of the Programming verifier stage within the PhysVEC framework. (a) The evaluated efficiency corresponding to input tokens ($T_i$), output tokens ($T_o$), and consecutive tool utilization ($T_c$) for LLMs across four topics. Horizontal dashed lines denote the average performance of the $4$ models within each respective topic. Comprehensively, these metrics identify \texttt{nqs} as the most challenging task domain within QMP-Bench, while \texttt{Claude Sonnet 4} consistently exhibits the best overall performance. (b) The marginal utility of tool usage (primarily document retrieval from the local knowledge base) contrasting PhysVEC with the ReAct-RAG baseline. The dashed lines depict the baseline marginal benefit of employing retrieval tools within ReAct-RAG. Solid circles situated above these dashed lines explicitly demonstrate the markedly higher tool-use efficiency intrinsic to PhysVEC. (c) Progression of the pass rate evaluated via repeated trials on $8$ initially failed \texttt{nqs} tasks using \texttt{Gemini 2.5 Flash}. The pass rate is defined as the fraction of tasks that succeed at least once within the first $N$ independent trials. This temporal dynamic clearly indicates that the PhysVEC framework exhibits a characteristic inference-time scaling effect.}
    \label{fig:efficiency}
\end{figure*}

Scientific simulations using domain-specific libraries are frequently accompanied by subtle errors in programming, such as third-party library calls, interface compatibility, and data structures. PhysVEC addresses this challenge by performing a global and systematic diagnosis and error correction across the script. This framework effectively suppresses LLMs' hallucinations and improves programming correctness and overall executability. We evaluated several LLMs on our benchmark dataset, including \texttt{GPT-5.1}, \texttt{Gemini 2.5 Flash}, \texttt{Qwen3-Max}, and \texttt{Claude Sonnet 4}. PhysVEC substantially improves the executability of generated scripts relative to baselines across the full QMP-Bench dataset.

In this stage, we compared the results of the PhysVEC framework with three baselines. In addition to former-mentioned ReAct and ReAct-RAG, we also include: 
\begin{itemize}
\item \texttt{PhysVEC-1-shot}: corresponds to the raw output of the Author agent in our system, without any subsequent verification-and–correction iterations.
\end{itemize}

For each pair of LLMs and topics, we define executability $S$ as the fraction of scripts that run without errors. The results are shown in Fig.~\ref{fig:programtest} (a). As for PhysVEC (the maximum number of iterations is set to 8), after iterative refinement with the Programming verifier, all scripts become executable in most cases, yielding substantially higher $S$ than all baselines. Besides executability, in PhysVEC the unit tests and integration tests provide a finer-grained characterization of syntax correctness than a single binary success signal. We introduce an accuracy metric for these tests. In the unit test, accuracy measures the fraction of element functions that are validly defined and executable in isolation. In the integration test, accuracy indicates how far the execution has progressed along the level hierarchy. Figs.~\ref{fig:programtest} (b) and (c) compare the accuracy of unit tests and integration tests before (hollow markers, PhysVEC-1-shot) and after (solid markers, PhysVEC) iterative verification and error correction. Across all models and topics, the accuracy increases markedly, indicating that more individual functions become executable and longer execution chains are achieved. These trends provide additional evidence for the effectiveness of the verification and error correction design in our framework.

Beyond the final performance, Fig.~\ref{fig:programtest} (d) illustrates the error correction progress by plotting executability against the number of iterations. We select the trajectories of \texttt{GPT-5.1} on the \texttt{dmrg} and \texttt{nqs} topics as representative examples, noting that all other model-topic combinations exhibit similar behaviors. From these curves, two specific conclusions can be drawn: First, in contrast to the sluggish growth of the baselines, PhysVEC achieves a much steeper initial climb. Second, the PhysVEC framework systematically converges to a substantially higher executability saturation value. These distinct contrasts firmly validate the PhysVEC's robust error-resolving capability driven by the explicitly designed self-verification mechanisms for programming correctness.

We attribute this performance gain to the framework's ability to reveal all errors simultaneously and apply parallel error corrections at each iteration. More specifically: (1) To analyze and compare with PhysVEC-1-shot, we consider the structured script to have a simple probabilistic error model: if each element function is correct with probability upper bounded by $p$, then the overall success probability of the script decays exponentially according to $P_S=p^{N_E}$ ($N_E$ denotes the number of element functions in a script). The PhysVEC error-correction mechanism on each element function effectively counteracts this exponentially shrinking success probability (as $N_E$ grows), pushing the executability towards $1$. (2) Compared to ReAct and ReAct-RAG baselines, the Programming verifier conducts comprehensive unit tests that examine all element function definitions simultaneously, surfacing all programming issues in a single round. By contrast, conventional LLM iteration loops can only detect and patch the first runtime error encountered in each execution. (3) Compared to all baselines, the integration tests in PhysVEC both uncover compatibility problems between different blocks (such as mismatched interfaces or inconsistent data structures) and cross‑validate the results of unit tests, jointly improving the robustness and efficiency of this framework.

We also investigated the efficiency of token and tool use of PhysVEC and baselines, as reported in Fig.~\ref{fig:efficiency} (a) and (b). For each task, we tracked the total input tokens $T_i$, output tokens $T_o$, and number of tool calls $T_c$ (primarily for retrieving documentation or manuals). Fig.~\ref{fig:efficiency} (a) compares the efficiency of PhysVEC across models and topics, showing that tasks in \texttt{nqs} are generally more challenging, while \texttt{dmrg} and \texttt{dft} are relatively easier. In general, \texttt{Claude Sonnet 4} achieves the best overall performance in the programming test and uses tokens and tools most efficiently.

We compare the marginal utility of tool calls between the PhysVEC framework and the ReAct-RAG baseline, as shown in Fig.~\ref{fig:efficiency} (b). Compared to ReAct, ReAct-RAG retrieves library documentation and manuals during the repair loop, thereby achieving higher executability. In Fig.~\ref{fig:efficiency} (b), crosses denote the results of ReAct (with the point at $T_c=0$) and ReAct-RAG (with the point at $T_c>0$), and the dashed line indicates the gain in executability attributable to the retrieval calls in ReAct-RAG. In the figure, solid circles represent the PhysVEC results. In most model-topic pairs, PhysVEC consistently lies above the dashed lines, indicating that it makes more effective use of retrieval calls than the ReAct-RAG baseline. In some cases, tasks are relatively easy and the baseline executability is already close to $1$, so the marginal gains saturate.

As shown in Fig.~\ref{fig:efficiency} (c), we also conducted inference-time scaling experiments \cite{snell2025scaling} on the failed tasks in \texttt{nqs} for \texttt{Gemini-2.5-Flash}. In the figure, we present the increase of the total pass rate, defined as the fraction of these tasks that pass at least once within the first $N$ repeated trials. This result suggests that PhysVEC, with verification and error correction, exhibits an inference-time scaling effect: increasing the number of repeated trials improves performance on most tasks.

\subsection{Scientific verification and error correction}

\begin{figure*}
    \centering
    \includegraphics[width=0.75\linewidth]{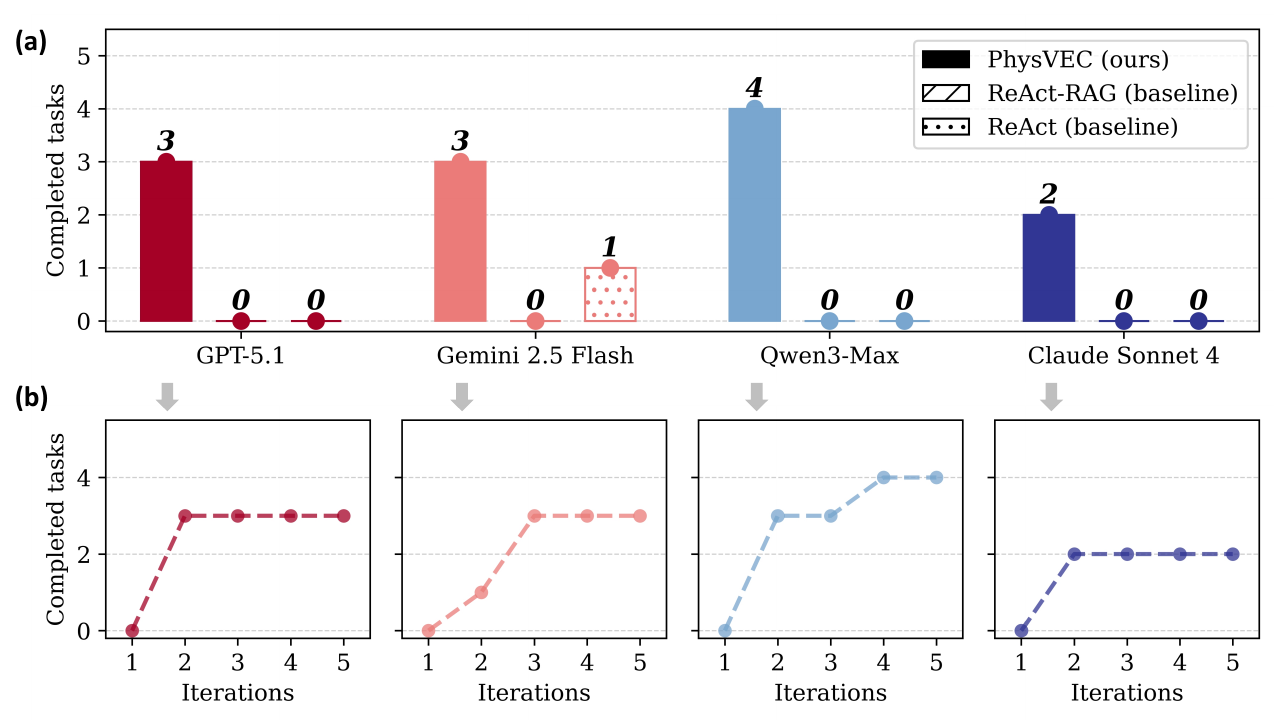}
    \caption{Enhancements in physical validity achieved by the PhysVEC framework on the QMP-Bench subset (consisting of $5$ \texttt{dmrg} tasks). (a) Comparison of the scientific test performance between PhysVEC and conventional baselines (ReAct-RAG and ReAct). Notably, the PhysVEC results represent the final performance after both the Programming and Scientific verifiers, whereas the baselines lack these dual verification modules. The vertical axis denotes the absolute number of ``completed tasks" that successfully yield physically valid results. The results explicitly demonstrate that PhysVEC significantly outperforms the baselines across all evaluated LLMs, securing a substantially higher number of physically validated completions. (b) The cumulative number of completed tasks under the PhysVEC framework as a function of iteration steps of the Scientific verifier. The curves illustrate a steady rise in successfully completed tasks as the iterations progress.}
    \label{fig:scitest2}
\end{figure*}

In scientific simulations, it is not sufficient for scripts to be merely grammatically correct. In real research settings, they must also be scientifically valid, reflecting the underlying definitions, assumptions, and constraints of the target problem. Moreover, for AI-driven automated scientific research to be seriously adopted by the community, the system should also provide interpretable and human-auditable evidence supporting its results.

In the PhysVEC framework, the Scientific verifier takes the scripts that pass the programming test and further examines their physical validity. As illustrated in Fig.~\ref{fig:workflow} (a), the pipeline involves three main steps: the rubric test, the physical assertion test, and the convergence test. These three steps ensure successful reproduction of the target task while providing convincing evidence along with the final output.

In the rubric test, the Scientific verifier uses rubrics to check and refine the evaluated scripts. For each target task, a rubric is constructed to include the specification of the physical system, the computational method, and important physical/numerical parameters \cite{PaperBench}. Guided by the rubric, the Scientific verifier inspects the LLM-generated code for semantic correctness and proposes modifications when inconsistencies are detected. 

In the physical assertion test, the Scientific verifier checks whether the script's execution results satisfy predefined physical constraints. In conventional numerical studies, human researchers routinely validate new codes by testing them under carefully chosen regimes or conditions where the behavior or the exact answer is known. PhysVEC adopts the same philosophy. For each task, we predefine three types of physical assertions: (1) settings where exact diagonalization is feasible at small system sizes (limiting case tests), (2) settings whose outcomes are constrained by known symmetries (symmetry tests), and (3) parameter regimes with established analytical results (analytical tests). The Scientific verifier modifies and executes the script to compare the outputs against the predefined answers (see Appendix~\ref{sec:append-scitest} for details).

In the convergence test, the Scientific verifier adjusts the calculation parameters to ensure numerical convergence has been achieved. In scientific simulations, numerical parameters often affect both the validity and the precision of the results. Through this step, the Scientific verifier completes the workflow to ensure sound outcomes.

\begin{figure*}[htbp]
    \centering
    \includegraphics[width=0.85\textwidth]{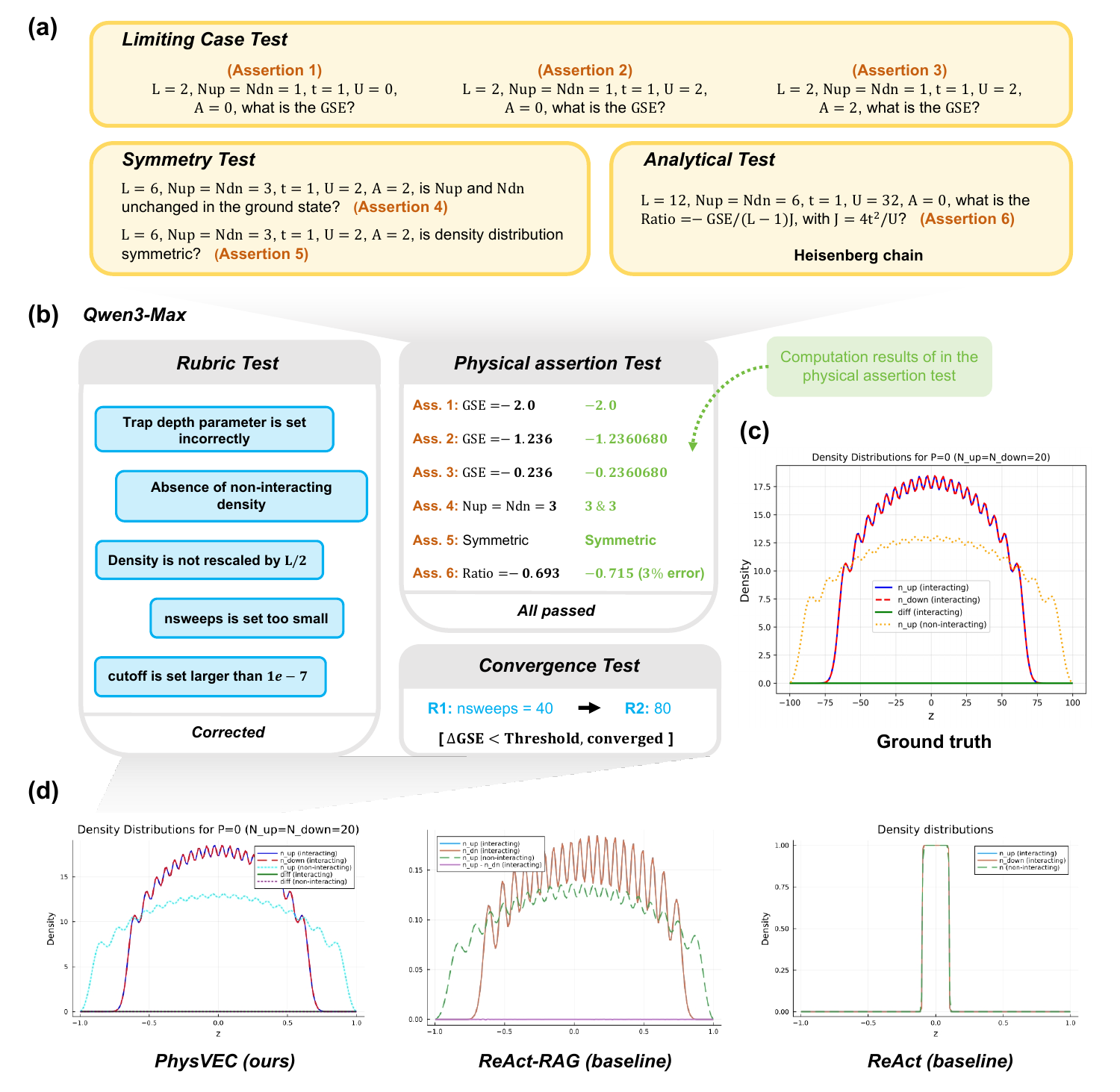}
    \caption{Case study demonstrating how PhysVEC elevates physical validity and yields human-auditable evidence in the Scientific verifier. The target task is reproducing the density distributions of interacting electrons in a harmonic trap \cite{PhysRevLett.100.110403}. (a) Physical assertions explicitly designed for this task. These assertions are concisely categorized into three rigorous checks: small-size limiting cases tractable by exact diagonalization, constraints dictated by established symmetries, and parameter regimes with known analytical solutions. (b) Detailed verification and refinement workflow executed by \texttt{Qwen3-Max}. Driven by rubrics, physical assertions, and convergence checks, the agent accurately pinpoints and corrects physical faults, including the incorrect trap shape configuration, the absent density rescale and the insufficient convergence parameter. The explicit records from these validation steps serve as human-auditable evidence, fundamentally ensuring the reliability of the final results. (c) The target ground-truth density distribution extracted from the original literature. (d) Comparison of the final results generated by PhysVEC and the baseline frameworks. The output of PhysVEC exhibits excellent agreement with the ground truth, marking a successful reproduction. In contrast, both the baseline methods fail: ReAct-RAG produces erroneous distributions due to unconverged DMRG computations, while the standard ReAct workflow misconfigures an excessively deep harmonic trap, which distorts the physical state. }
    \label{fig:scittest}
\end{figure*}

Due to computational constraints, we conducted the scientific test on a subset of QMP-Bench consisting of five tasks. Fig.~\ref{fig:scitest2} presents the scientific test results of four models under the PhysVEC framework, compared to the direct execution of scripts generated by the baselines. The execution results are compared to the ground truth, and the number of successfully completed tasks is reported in Fig. 5 (a). PhysVEC performs substantially better than both baselines, demonstrating the Scientific verifier's ability to improve the physical validity of the scripts and ensure accurate research results. Fig.~\ref{fig:scitest2} (b) shows the cumulative number of tasks completed as a function of scientific test iteration for each model. Within the PhysVEC framework, the Scientific verifier iteratively refines the LLM-generated script via rubric tests and physical assertion tests until no physical errors are detected. As expected, more tasks are successfully completed as iterations proceed. It is worth noting that, despite being evaluated on a subset rather than the entire QMP-Bench, the consistent improvements observed across all LLMs here clearly highlight the essential role of the Scientific verifier and its design of physics-informed verifications. 

As a concrete case study, we focused on producing Fig.~1 (a) of Tezuka et al. \cite{PhysRevLett.100.110403}, which depict the ground-state density distributions of interacting electrons confined in a one-dimensional harmonic trap. The procedure (driven by \texttt{Qwen3-Max}) is shown in Fig.~\ref{fig:scittest}. In this case study, we constructed $6$ physical assertions, as shown in Fig.~\ref{fig:scittest} (a). In the PhysVEC pipeline, the Scientific verifier identified several error issues in the rubric test, including an incorrect definition of the harmonic-trap depth and the omission of the rescaling for the computed density, as shown in the left panel of Fig.~\ref{fig:scittest} (b). Then the Scientific verifier faithfully adapted the original code according to the assertions, and the results successfully passed all the assertion checks, as shown in the middle panel of Fig.~\ref{fig:scittest} (b). Finally, the Scientific verifier increased \texttt{nsweeps} from $40$ to $80$ and monitored the change in the calculated ground-state energy. It determined that the simulation had converged and output the results. In Fig.~\ref{fig:scittest} (d), we compare the final results of the PhysVEC and other baselines. The figure from PhysVEC is almost identical to the ground truth in Fig.~\ref{fig:scittest} (c). In the ReAct‑RAG result, the density distribution of the interacting electrons exhibits large oscillations and is not centered in the harmonic trap, indicating that the calculation has not converged. In the ReAct result, the electrons are all concentrated in a small region, which is caused by an incorrect harmonic trap depth setting in the code.

In the PhysVEC, the Scientific verifier performs effectively because it enforces a comprehensive verification suite that jointly ensures that LLM-generated scripts (1) faithfully instantiate the intended physical system and (2) execute numerically reliable computations. From an information-theoretic perspective, rubrics and physical assertions act as explicit constraints that narrow the space of plausible solutions: if the candidate space is $\mathcal{H}$ with uncertainty $H_0=\log|\mathcal{H}|$, and each test contributes an information gain $I_i$, then the remaining uncertainty decreases as $H_{\mathrm{final}}=H_0-\sum_i I_i$. This implies that the number of candidate solutions that satisfy all tests decreases exponentially with the accumulated information gain, scaling as $\exp\!\left(-\sum_i I_i\right)$. In contrast, baselines solely depend on the LLM's intrinsic task-understanding capability and its implicit physics knowledge, which can be inadequate. 

\section{Discussion}

In this work, we addressed a critical gap in AI-driven scientific research: the absence of systematic verification and correction mechanisms for LLM-generated results at both the programming and scientific levels. We introduced PhysVEC, a multi-agent AI physicist framework that moves beyond conventional ReAct paradigms by comprehensively diagnosing failures at each stage and applying error corrections accordingly. By enforcing a structured script, PhysVEC conducts unit tests, integration tests, rubric tests, physical assertion tests, and convergence tests throughout the pipeline. Leveraging the verification suite described above, PhysVEC achieves automated error correction that substantially improves both the executability and physical validity of the scripts, while providing interpretable evidence.

We curated QMP-Bench, to our knowledge the first end-to-end research-level quantum many-body physics benchmark dataset comprising $100$ tasks drawn directly from $21$ high impact articles. Unlike prior benchmarks based on human-curated or pre-digested problems, QMP-Bench is grounded in original research articles and requires challenging simulations in realistic research settings. We evaluated four frontier LLMs (\texttt{GPT-5.1}, \texttt{Gemini 2.5 Flash}, \texttt{Qwen3-Max}, and \texttt{Claude Sonnet 4}) on QMP-Bench. PhysVEC with verification and error-correction design produces faithful and interpretable physical results, outperforming all the baseline methods.

There are several important open questions for future research. First, scientific tests still rely on human expertise. Automatically synthesizing rubrics and physical assertions is an important next step. Second, current LLMs still lack sufficient domain expertise and self-reflection, and may fail to correct underlying errors even after multiple verification iterations, such as subtle inaccuracies in Hamiltonian constructions. Third, an AI physicist should be capable of autonomously generating novel physical hypotheses and proposing reasonable predictions, moving toward truly end-to-end autonomous scientific discovery. The further development of PhysVEC will provide a principled and scalable framework for autonomous, reliable, and interpretable scientific discovery.

\begin{acknowledgments}
DL acknowledges support from Beijing Municipal Science and Technology Commission and Zhongguancun Science Park Administrative Committee (No. 20251090054). JH acknowledges support from the Natural Science Foundation of Jiangsu Province (No. BK20250404), the Youth Science and Technology Talent Support Project of Jiangsu Province (No. JSTJ-2025-600). 
\end{acknowledgments}

\clearpage
\appendix

\section{\label{appendixsec-packages}Computational methods and numerical packages}

In the QMP-Bench dataset and the PhysVEC framework, we incorporate four major computational paradigms. For each domain, the agents are equipped with specific and widely adopted numerical packages/softwares to construct and execute simulations:

\begin{itemize}
    \item \textbf{Tensor network:} Tensor network methods represent quantum states as networks of low-rank tensors, providing a compact description of the entanglement structure in many-body systems. In this work, our agents are equipped with the ITensors package \cite{ITensors} in Julia to implement tensor-network-based computations.
    
    \item \textbf{Neural network ansatz:} Quantum many-body wavefunctions can also be approximated by neural network architectures, which provides a flexible variational ansatz that can efficiently capture complex correlations and entanglement patterns combined with various variational optimization methods. In this work, our agents employ the NetKet package \cite{NetKet,NetKet3} to construct neural-network wavefunctions and perform the corresponding variational calculations.
    
    \item \textbf{Quantum circuit simulation:} Quantum many-body systems can also be simulated within the quantum circuit framework. Algorithms such as Trotterized time evolution, Iterative Quantum Phase Estimation (IQPE), and Variational Quantum Eigensolvers (VQE) are frequently used for various systems. In this work, the agents use the Qiskit package \cite{Qiskit} to simulate the quantum circuit and the corresponding algorithms on a classical computer.
    
    \item \textbf{Density-functional theory calculation:} Density-functional theory (DFT) uses the particle density to represent many-electron systems and is widely used for calculating electronic structure and spectral properties. In this work, the agents use the ORCA \cite{ORCA} to perform DFT calculations and analyze the corresponding properties. In ORCA, integration tests are not applicable because the program does not accept incomplete input files.
\end{itemize}

\section{\label{appendixsec-reference}Local knowledge base and reference document}
In the PhysVEC framework, we prepared a local knowledge base that includes the official GitHub repositories of numerical packages and the user manuals of the corresponding software. The Author agent consults this base to generate scripts with syntax and API calls for professional numerical libraries/softwares. After the Programming verifier identifies errors in the Author-agent-generated scripts through unit tests and integration tests, it also queries this knowledge base for each error to obtain the information needed for correction.

\section{\label{appendixsec-verifierlib}Unit test verifier library}

\begin{figure*}[htbp]
    \centering
    \includegraphics[width=0.75\textwidth]{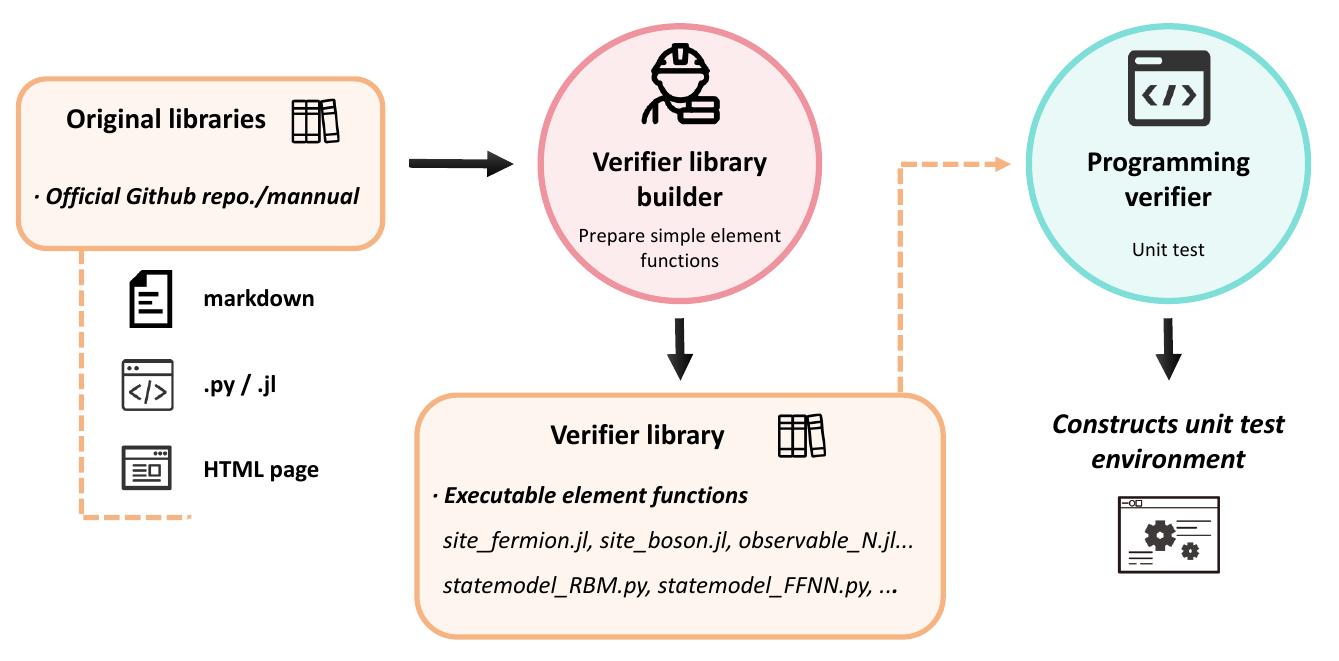}
    \caption{Schematic of the Verifier library builder, which distills official repositories/manuals into a verifier library that serves as reference material for unit tests.}
    \label{fig:verifier-library}
\end{figure*}

We defined a Verifier library builder to prepare the verifier library (or background code base) required for unit tests, as shown in Fig.~\ref{fig:verifier-library}. During unit tests, the Programming verifier selects appropriate functions from this verifier library to instantiate the test environment for the target element function. The background functions and the function to be tested form a complete computational pipeline. The Verifier library builder operates independently of the Programming verifier. It directly reads the official repositories or manuals of numerical libraries and generates a collection of potentially reusable element functions. Through extensive refine-and-retrieve cycles, it ultimately produces syntactically well-defined executable functions and stores them in the verifier library.

\section{\label{sec:append-scitest}Rubric test and physical assertion test}

\begin{figure*}[htbp]
    \centering
    \includegraphics[width=0.85\textwidth]{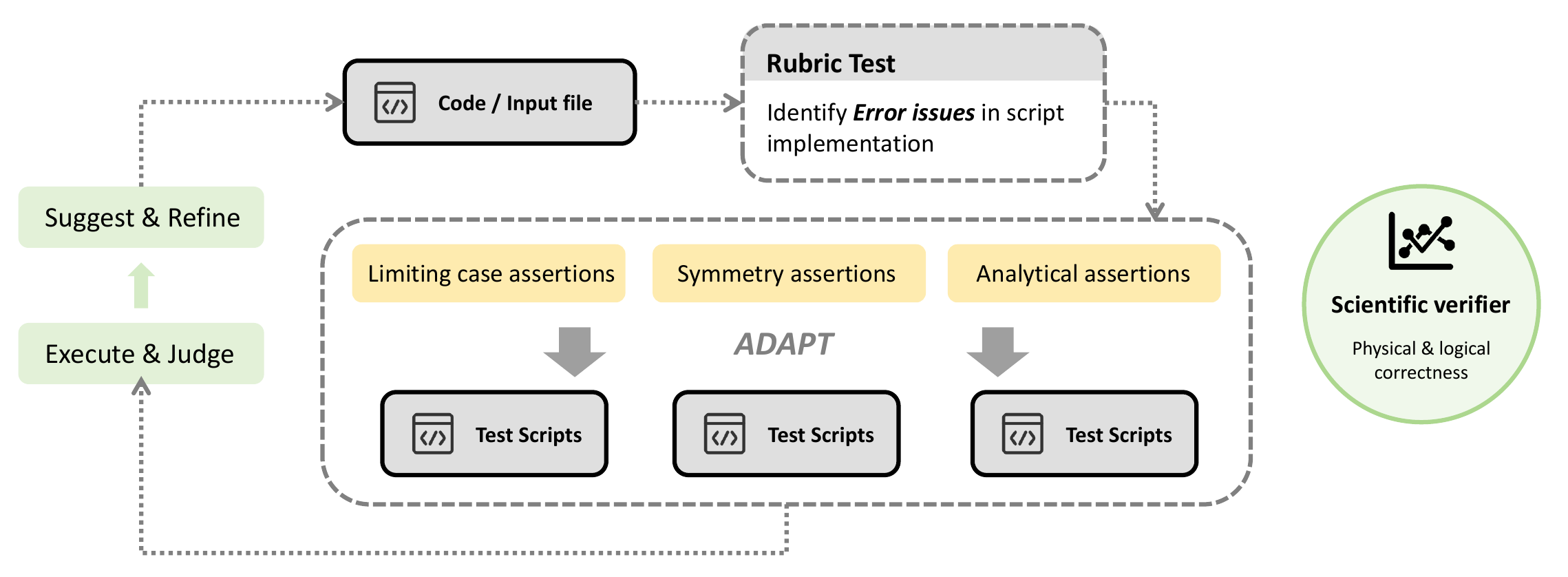}
    \caption{Schematic of the rubric test and the physical assertion test by Scientific verifier. }
    \label{fig:physical-assertion-demo}
\end{figure*}

The Scientific verifier iteratively refines the scripts produced by the Programming verifier through rubric tests and physical assertion tests to verify and ensure their physical and numerical‑setup validity, as shown in Fig.~\ref{fig:physical-assertion-demo}. 

In the rubric test, the agent identifies all error issues in the script that do not satisfy the predefined requirements. In practice, PhysVEC repeats this procedure multiple times, increasing the LLM temperature across iterations to promote reasoning diversity, which improves the agent's decision accuracy and reduces erroneous judgments.

In the physical assertion test, the agent adapts the original script according to different assertions and runs the corresponding tests. The agent then aggregates the results to apply modifications to the script accordingly. The Scientific verifier continues this iterative loop until it reaches the predefined maximum iteration number or passes both tests simultaneously.

\section{\label{appendixsec-tokenuse}Token and tool usage statistics}

\begin{figure*}[htbp]
    \centering
    \includegraphics[width=0.8\textwidth]{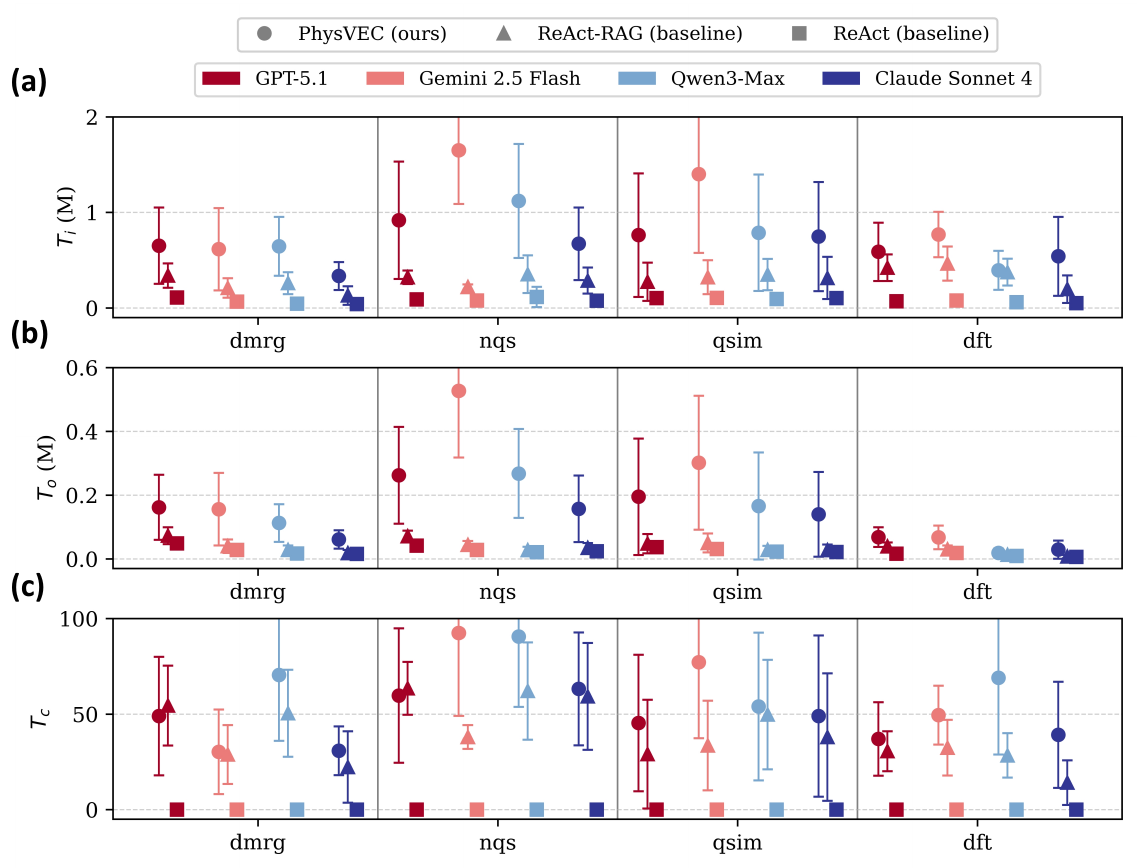}
    \caption{Token usage and tool calls in programming tests. (a) Input tokens. (b) Output tokens. (c) Tool calls. }
    \label{fig:usage}
\end{figure*}

Figs.~\ref{fig:usage} (a) and (b) report the token consumption of PhysVEC and the baselines in the programming test, averaged over all tasks for a given model and topic. As expected, our framework uses substantially more tokens than the baselines, due to the more complex multi‑agent architecture and repeated verification operations. However, the input token cost typically remains below $1$ million, while the output token cost is mostly under $0.3$ million, corresponding to only a few dollars per task on average. This cost is practically acceptable for automatic reproduction of scientific codes. Fig.~\ref{fig:usage} (c) shows the number of tool calls (mainly retrieve operations) in the tests. 

\begin{figure*}[htbp]
    \centering
    \includegraphics[width=0.6\textwidth]{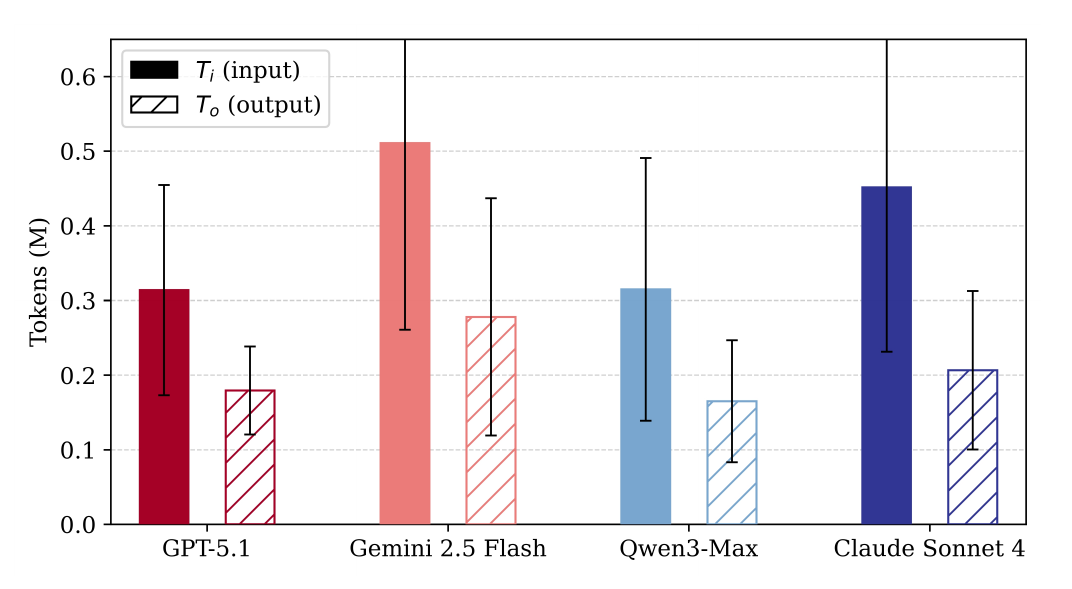}
    \caption{Token usage of the Scientific verifier in the PhysVEC framework. Solid bars represent input tokens and hatched bars represent output tokens.}
    \label{fig:scitest_usage}
\end{figure*}

Fig.~\ref{fig:scitest_usage} reports the token consumption of PhysVEC during the scientific test stage. Both input and output token costs remain mostly below $0.5$ million per task.

\section{\label{appendixsec-articlelist}Dataset article list}

To ensure transparency and provide a detailed overview of our benchmark dataset, we list the source articles partitioned by their underlying numerical methodologies below. The compiled literature spans a diverse range of quantum many-body simulations, encompassing tensor network algorithms (labeled \texttt{dmrg}, Table~\ref{dataset-dmrg}), neural network ansatz for quantum states (\texttt{nqs}, Table~\ref{dataset-nqs}), quantum circuit simulations for many-body systems (\texttt{qsim}, Table~\ref{dataset-qsim}), and density-functional theory calculations (\texttt{dft}, Table~\ref{dataset-dft}). Alongside the article information, we quantitatively breakdown the dataset by providing the number of selected tasks (figures/tables for reproduction) extracted from each article, thereby reflecting the comprehensive scope of our analysis.

\begin{table*}
\caption{\label{dataset-dmrg}Article list for \texttt{dmrg} tasks}
\begin{ruledtabular}
\begin{tabular}{ccc}
Journal & Title & Number of tasks \\ \hline

    Physical Review Letters     & \parbox{10cm}{\vspace{0.1cm}Phase diagram of spin-1 bosons on one-dimensional lattices\vspace{0.1cm}} & 5 \\
    Physical Review Letters     & \parbox{10cm}{\vspace{0.1cm}Correlations in a two--chain Hubbard model\vspace{0.1cm}} & 5 \\
    Physical Review Letters     & \parbox{10cm}{\vspace{0.1cm}Disorder induced quantum phase transition in random-exchange spin-1/2 chains\vspace{0.1cm}} & 6 \\
    Physical Review Letters     & \parbox{10cm}{\vspace{0.1cm}Plaquette ordered phase and quantum phase diagram in the spin-1/2 J1-J2 square Heisenberg model\vspace{0.1cm}} & 3 \\
    Physical Review Letters     & \parbox{10cm}{\vspace{0.1cm}Hidden order in one dimensional Bose insulators\vspace{0.1cm}} & 2 \\
    Physical Review Letters     & \parbox{10cm}{\vspace{0.1cm}Topological edge states in the one-dimensional super-lattice Bose-Hubbard model\vspace{0.1cm}} & 5 \\
    Physical Review Letters     & \parbox{10cm}{\vspace{0.1cm}Density-matrix renormalization group study of trapped imbalanced Fermi condensates\vspace{0.1cm}} & 5 \\
    Physical Review Letters     & \parbox{10cm}{\vspace{0.1cm}One hole in the two-leg t-J ladder and adiabatic continuity to the non-interacting limit\vspace{0.1cm}} & 4 \\
    Physical Review Letters     & \parbox{10cm}{\vspace{0.1cm}Evidence for a superfluid density in t–J ladders\vspace{0.1cm}} & 1 \\

\end{tabular}
\end{ruledtabular}
\end{table*}

\begin{table*}
\caption{\label{dataset-nqs}Article list for \texttt{nqs} tasks}
\begin{ruledtabular}
\begin{tabular}{ccc}
Journal & Title & Number of tasks \\ \hline

    Physical Review Letters     & \parbox{10cm}{\vspace{0.1cm}Symmetries and many-body excited states with neural-network quantum states\vspace{0.1cm}} & 4 \\
    Physical Review X     & \parbox{10cm}{\vspace{0.1cm}Quantum entanglement in neural network states\vspace{0.1cm}} & 4 \\
    Physical Review B     & \parbox{10cm}{\vspace{0.1cm}Quantum skyrmion dynamics studied by neural network quantum states\vspace{0.1cm}} & 3 \\
    Science & \parbox{10cm}{\vspace{0.1cm}Solving the quantum many-body problem with artificial neural networks\vspace{0.1cm}} & 7 \\
    
\end{tabular}
\end{ruledtabular}
\end{table*}

\begin{table*}
\caption{\label{dataset-qsim}Article list for \texttt{qsim} tasks}
\begin{ruledtabular}
\begin{tabular}{ccc}
Journal & Title & Number of tasks \\ \hline

    npj Quantum Information & \parbox{10cm}{\vspace{0.1cm}Simulating quantum many-body dynamics on a current digital quantum computer\vspace{0.1cm}} & 9 \\
    EPJ Quantum Technology  & \parbox{10cm}{\vspace{0.1cm}Quantum simulation of the Hubbard model on a graphene hexagon: Strengths of IQPE and noise constraints\vspace{0.1cm}} & 5 \\
    Nature communications   & \parbox{10cm}{\vspace{0.1cm}Observing ground-state properties of the Fermi-Hubbard model using a scalable algorithm on a quantum computer\vspace{0.1cm}} & 6 \\
    
\end{tabular}
\end{ruledtabular}
\end{table*}

\begin{table*}
\caption{\label{dataset-dft}Article list for \texttt{dft} tasks}
\begin{ruledtabular}
\begin{tabular}{ccc}
Journal & Title & Number of tasks \\ \hline

    JACS & \parbox{10cm}{\vspace{0.1cm}Spectroscopic evidence for a 3d10 ground state electronic configuration and ligand field inversion in [Cu(CF3)4]1–\vspace{0.1cm}} & 6 \\
    JACS & \parbox{10cm}{\vspace{0.1cm}Nature of S-states in the Oxygen-evolving complex resolved by high-energy resolution fluorescence detected X-ray absorption spectroscopy\vspace{0.1cm}} & 6 \\
    JACS & \parbox{10cm}{\vspace{0.1cm}Combining valence-to-core X-ray emission and Cu K-edge X-ray absorption spectroscopies to experimentally assess Oxidation state in organometallic Cu(I)/(II)/(III) complexes\vspace{0.1cm}} & 5 \\
    Nature communications & \parbox{10cm}{\vspace{0.1cm}Quantitative prediction of rate constants and its application to organic emitters\vspace{0.1cm}} & 4 \\
    JACS & \parbox{10cm}{\vspace{0.1cm}Valence-to-core X-ray emission spectroscopy: A sensitive probe of the nature of a bound ligand\vspace{0.1cm}} & 5 \\
    
\end{tabular}
\end{ruledtabular}
\end{table*}

\nocite{*}

\clearpage
\bibliography{refs_main}

\clearpage
\newpage
\section*{Supplemental Material for ``Towards Verifiable and Self-Correcting AI Physicists \\for Quantum Many-Body Simulations''}

\subsection*{Pipeline of the PhysVEC Framework}
In this section, we detail the complete operational pipeline of the PhysVEC framework.

\subsubsection*{Author agent}
\textbf{Preparation. }
We initiate the pipeline by feeding the target task description into the Author agent, as exemplified by the following JSON file: 

\begin{lstlisting}
{
   "pdf_path": "sources",
   "pdf_name": "PhysRevLett.100.110403",
   "subplot_name": "Fig._1_a_right",
   "User_requests": "Only focus on the reproduction of Fig. 1a-right panel (density distribution for P=0). The caption of the figure containing this subplot begins with 'On-site pair correlation xxx (left column)  of spin-up and spin-down ...'. You ONLY reproduce the density distributions (spin-up, spin-down, difference of these two, and non-interacting) of the (20, 20) case. You MUST use ITensors and ITensorMPS packages. ",
   "rubrics_name": "PhysRevLett.100.110403_Fig._1_a_right_rubrics.json",
   "LimitingCases": [
      ... (More details later)
   ],
   "ConvergenceCases": [
      ... (More details later)
   ]
}
\end{lstlisting}

The pdf\_path and pdf\_name fields locate the source article, while the User\_requests field identifies the target figure and applies specific constraints, such as reproducing only specific subplots or focusing on distinct parameter configurations.

Upon processing the original article, the Author agent generates free-format reasoning and a plan, as illustrated below:

\begin{lstlisting}
{
   "plan": "The goal is to reproduce the right panel of Fig. 1(a), ...\n\n**1. System Definition and Hamiltonian:**\n   The system is a 1D attractive Hubbard model with a harmonic on-site potential. ...\n\n**2. Parameters from the paper:**\n   - Lattice size (L): L = 200 sites.\n   - Number of spin-up atoms (N_up): N_up = 20.\n ...\n\n**3. DMRG Calculation Steps:**\n   a. **Initialize MPS:** Create an initial Matrix Product State (MPS) ..."
}
\end{lstlisting}

Subsequently, guided by the plan, the Author agent invokes the retrieve-mcp tool to query the knowledge base for relevant information, such as functions in the ITensors and ITensorMPS packages, which is then concentrated into a query\_summary file:

\begin{lstlisting}
{
  "summary": [
    {
      "query": "AutoMPO creation annihilation operators Hubbard model",
      "answer": "To create creation and annihilation operators for a Hubbard model using `AutoMPO` (or `OpSum`), you first define `\"Electron\"` site types. Then, you can use ..."
    },
    {
      "query": "DMRG algorithm parameters maxdim cutoff nsweeps",
      "answer": "The DMRG algorithm in ITensors.jl and ITensorMPS.jl uses several key parameters to control its execution and accuracy:\n\n- `nsweeps`: The number of sweeps (iterations) the DMRG algorithm will perform. Each sweep ..."
    },
    ...
  ]
}
\end{lstlisting}

\textbf{Script generation. }
Integrating the information from the plan and the query\_summary, the Author agent generates structured script that reproduces the target problem:

\begin{lstlisting}
using ITensors
using ITensorMPS
using Plots
using Random

function site_Fermion_Hubbard(L::Int)::Vector{<:Index}
    sites = siteinds("Fermion", L; conserve_qns=true)
    return sites
end

function hamiltonian_Hubbard_potential(sites::Vector{<:Index}, ...)::MPO
    ampo = OpSum()
    ...
    H = MPO(ampo, sites)
    return H
end

function initialstate_random_product(sites::Vector{<:Index}, ...)::MPS
    ...
    psi0 = MPS(sites, state)
    return psi0
end

function observable_density_distributions(psi::MPS, L::Int)::Tuple{Vector{Float64}, Vector{Float64}}
    ...
    return n_up, n_down
end

function effector_dmrg_ground_state(H::MPO, psi0::MPS, ...)::Tuple{Float64, MPS}
    energy, psi = dmrg(H, psi0; nsweeps=nsweeps, maxdim=maxdim, cutoff=cutoff)
    return energy, psi
end

function run_density_calculation(L::Int, N_up::Int, ...)
    sites = site_Fermion_Hubbard(L)
    psi0 = initialstate_random_product(sites, L, N_up, N_down)

    # Interacting case
    H_interacting = hamiltonian_Hubbard_potential(sites, L, t, U, V)
    energy_interacting, psi_interacting = effector_dmrg_ground_state(H_interacting, psi0, nsweeps, maxdim, cutoff)
    n_up_interacting, n_down_interacting = observable_density_distributions(psi_interacting, L)

    # Non-interacting case (U=0)
    ...

    return n_up_interacting, n_down_interacting, n_up_nonint, n_down_nonint
end

...

function main()
    # --- Parameters from the paper ---
    ...

    # --- DMRG parameters ---
    ...

    # --- Data collection ---
    ...
end

main()
\end{lstlisting}

As noted in the main text, the generated script follows a rigorous structural design, initially defining discrete element functions (e.g. "site\_xxx", "hamiltonian\_xxx", "effector\_xxx", etc.) before subsequently orchestrating these functions for the final data calculation (e.g. "run\_density\_calculation"). This rigid script structure serves as the foundation for subsequent efficient verification and error correction.

\subsubsection*{Programming verifier}
The Programming verifier ensures the programming correctness of the script through an iterative loop. In each iteration, it conducts unit tests and integration tests on the script, followed by targeted retrieval and correction to address any identified issues.

\textbf{Unit test. }
In the unit test stage, the Programming verifier evaluates the definition of every element function individually. To establish a fully functional testing environment, it selects relevant background functions from the Verifier library (a repository containing amounts of verified definitions, more details in the main text) according to the specific function under test, as demonstrated below:

\begin{lstlisting}
using ITensors, ITensorMPS

# ==== Background functions from the Verifier library ====
# site
function site_verifier_electron(N::Int; conserve_qns::Bool=false, conserve_sz::Bool=false, conserve_nf::Bool=false, conserve_nfparity::Bool=false)
    sites = siteinds(...)
    return sites
end

# initialstate
function initialstate_verifier_electron(sites, states; linkdims::Int=1)
    ...
        psi0 = MPS(sites, states)
    end
    return psi0
end

# effector
function effector_verifier_electron(H, psi0; nsweeps::Int=5, maxdim::Vector{Int}=[10, 20, 100], cutoff::Vector{Float64}=[1e-10])
    energy, psi = dmrg(H, psi0; nsweeps=nsweeps, maxdim=maxdim, cutoff=cutoff)
    return energy, psi
end

# observable
function observable_verifier_electron(psi::MPS, sites::Vector{<:Index}; ops::Vector{String}=["Ntot", "Sz"])
    ...
    results["correlation_matrix_up"] = correlation_matrix(psi, "Cdagup", "Cup")
    return results
end

# ==== Element function under test ====
function hamiltonian_Hubbard_potential(sites::Vector{<:Index}, L::Int, t::Float64, U::Float64, V::Float64)::MPO
    ampo = OpSum()
    ... (definition from the script from Author agent)
    H = MPO(ampo, sites)
    return H
end

# ==== Complete workflow/procedure ====
function main()
    # --- Parameters settings ---
    ...

    # 1. Create sites
    sites = site_verifier_electron(L; conserve_qns=true)
    
    # 2. Create initial state
    psi0 = initialstate_verifier_electron(sites, states; linkdims=10)
    
    # 3. Create Hamiltonian using the target function
    H = hamiltonian_Hubbard_potential(sites, L, t, U, V)
    
    # 4. Run DMRG
    energy, psi = effector_verifier_electron(H, psi0; nsweeps=nsweeps, maxdim=maxdim, cutoff=cutoff)
    
    # 5. Measure observables
    results = observable_verifier_electron(psi, sites; ops=["Nup", "Ndn", "Ntot", "Sz"])
end

main()
\end{lstlisting}

As demonstrated above, functions with ‘verifier’ in their names are sourced from the Verifier library. The Programming verifier concurrently constructs these environments for all element functions. Ultimately, it generates a comprehensive report that includes the correctness of each definition along with any possible error messages based on the execution results. A typical report looks like:

\begin{lstlisting}
"judge_result": [
{
  "target_file_name": "verify_effector_dmrg_ground_
         state_unittest.jl",
  "judge": "correct",
  "errors": "None"
},
...
{
  "target_file_name": "verify_observable_density_
         distributions_unittest.jl",
  "judge": "wrong",
  "errors": "The `observable_density_distributions` function attempts to call `expect(psi, \"Nup\", i)` and `expect(psi, \"Ndn\", i)`. However, the `ITensorMPS.expect` function ..."
}
]
\end{lstlisting}

\textbf{Integration test. }
In the integration test, the Programming verifier evaluates the interoperability among various element functions based on the script’s run function (generically named as "run\_xxx"), assessing the overall executability of the script. For example, for the original run function defined as: 

\begin{lstlisting}
function run_density_calculation(L::Int, N_up::Int, ...)
    sites = site_Fermion_Hubbard(L)
    psi0 = initialstate_random_product(sites, L, N_up, N_down)
    
    H_interacting = hamiltonian_Hubbard_potential(sites, L, t, U, V)
    energy_interacting, psi_interacting = effector_dmrg_ground_state(H_interacting, psi0, ...)
    n_up_interacting, n_down_interacting = observable_density_distributions(psi_interacting, L)
  
    return n_up_interacting, n_down_interacting, n_up_nonint, n_down_nonint
end
\end{lstlisting}

The Programming verifier decomposes the run function into a hierarchy of testing levels, starting from including only a single element function:

\begin{lstlisting}
function run_density_calculation(L::Int, N_up::Int, ...)
    sites = site_Fermion_Hubbard(L)
  
    return sites
end
\end{lstlisting}

to the complete assembly. At each incremental level, a new element function is introduced in its original sequential order. These scripts are run in parallel, yielding any possible compatibility issues within the original script. A report is then generated as:

\begin{lstlisting}
"run_functions": [
    {
      "run_function": "run_density_calculation",
      "error_analysis": "The integration test failed at level-2 because the `initialstate_random_product` function, which was newly introduced at this level, attempts to create an MPS using invalid state names for 'Fermion' sites ...",
      "correct_ratio": 0.125,
      "successful_element_functions": [
        "site_Fermion_Hubbard"
      ]
    }
]
\end{lstlisting}

In this example, the element functions for constructing sites and the initial state have both passed the unit test stage separately, but they prove to be incompatible when combined.

\textbf{Error correction. }
Following the unit and integration tests, the Programming verifier summarizes the reports from both phases to generate a repair suggestion, which itemizes the specific locations requiring modification. Subsequently, the Programming verifier addresses and resolves each issue sequentially. This "unit test - integration test - correction" loop is operated iteratively until the script can be executed without any programming error.

\subsubsection*{Scientific verifier}
Building upon the script that has passed the aforementioned programming tests, the Scientific verifier ensures its physical validity through an iterative loop. In each cycle, it conducts a rubric test and a physical assertion test. Once these tests are passed and the loop terminates, the Scientific verifier proceeds to a convergence test to make sure that the computational results are fully convergent.

\textbf{Rubric test. }
For the rubric test, the Scientific verifier checks the script’s content based on a human-curated scorecard, with a typical example shown below:

\begin{lstlisting}
{
  "weight": 1,
  1. "statement": "Code correctness",
      1.1 "issue": "The code plots the rescaled density distributions as a function of dimensionless coordinate ((i-i_{center})*2/L, where i is the site index).", 
      ...
          
  2. "statement": "Calculate the correctness of the model.", 
      2.1 "statement": "Correctness of constructing Hamiltonian", 
          2.1.1 "issue": "The code correctly constructs transition terms between all adjacent sites.", 
          ...
      ...
          
  3. "statement": "Correctness of calculation method.", 
      3.1 "statement": "Correctness of DMRG.", 
          3.1.1 "issue": "The code uses DMRG to obtain the ground states of the system.", 
          ...
      ...

  4. "statement": "Correctness of parameter settings.",
      4.1 "statement": "Correctness of calculation parameters", 
          4.1.1 "issue": "The code sets the cutoff parameter to be less than (or equal to) 1e-7, i.e. cutoff <= 1e-7.", 
          ...
      4.2 "statement": "Correctness of Model Parameter", 
          4.2.1 "issue": "The code sets the transition term coefficient to be$t=1$", 
          ...
}
\end{lstlisting}

Across all tasks, the rubric scorecard is constantly divided into four sections, inspecting the reproduction objective, physical system setup, computational method configuration, and parameter selection. The specific details within these sections are tailored to the characteristics of each task. The scorecard adopts a tree-like format where all leaf nodes are designated as ‘issues’. The Scientific verifier evaluates each issue to determine whether it is satisfied or not.

\textbf{Physical assertion test. }
During this phase, the Scientific verifier modifies the script according to assertion requirements formulated by human experts, executes it, and observes whether the output aligns with the expected answers. In general, we design physical assertions from three distinct aspects: limiting case tests, symmetry tests, and analytical tests. See more details in the 'Methods' of the main text. 

Typical physical assertions are presented as follows:

\begin{lstlisting}
{"content": "The ground state energy of this system with size L=2, Nup=Ndown=1, t=1, U=0 and A=0", 
"truth": "The ground state energy is -2. The relative error tolerance is +/- 0.05."
}
\end{lstlisting}

For the example above, the Scientific verifier modifies the entry point in the script from:

\begin{lstlisting}
function main()
    # --- Parameters from the paper ---
    L = 200  # Lattice size (scaled down from 200)
    N_up = 20 # Number of spin-up atoms (scaled down from 20)
    N_down = 20 # Number of spin-down atoms (scaled down from 20)
    t = 1.0 # Hopping parameter (energy unit)
    U = -4.0 # On-site interaction (scaled down, original U/t = -4)
    A = 0.16 # Potential depth (scaled down, original A/t = 0.16)
    V = 4 * A / L^2 # Harmonic potential strength
    U_nonint = 0.0 # Non-interacting on-site interaction

    # --- DMRG parameters ---
    nsweeps = 10 # Number of DMRG sweeps (scaled down from 10)
    maxdim = [100, 200, 200, 200, 200, 200, 200, 200, 200, 200] # Max bond dimensions (scaled down from [100, 200, ...])
    cutoff = [1E-6, 1E-7, 1E-8, 1E-9, 1E-10, 1E-10, 1E-10, 1E-10, 1E-10, 1E-10] # Truncation errors (scaled down from [1E-6, 1E-7, ...])

    z_coords, n_up_interacting, n_down_interacting, n_diff_interacting, n_diff_nonint = item_density_data(L, N_up, N_down, t, U, V, U_nonint, nsweeps, maxdim, cutoff)

    p = plot_density_distributions(z_coords, n_up_interacting, n_down_interacting, n_diff_interacting, n_diff_nonint, N_up, N_down)
    display(p)
end
\end{lstlisting}

to: 

\begin{lstlisting}
function main()
    # --- Limiting case parameters ---
    L = 2  # Lattice size
    N_up = 1 # Number of spin-up atoms
    N_down = 1 # Number of spin-down atoms
    t = 1.0 # Hopping parameter
    U = 0.0 # On-site interaction (non-interacting case)
    A = 0.0 # Potential depth (no potential)
    V = 4 * A / L^2 # Harmonic potential strength (will be 0)

    # --- DMRG parameters for a small system ---
    nsweeps = 10 # Number of DMRG sweeps
    maxdim = [10, 20] # Max bond dimensions (small for L=2)
    cutoff = [1E-8, 1E-10] # Truncation errors

    # Initialize sites and initial state
    sites = site_Fermion_Hubbard(L)
    psi0 = initialstate_random_product(sites, L, N_up, N_down)

    # Construct the Hamiltonian for the limiting case
    H = hamiltonian_Hubbard_potential(sites, L, t, U, V)

    # Calculate the ground state energy
    energy, psi = effector_dmrg_ground_state(H, psi0, nsweeps, maxdim, cutoff)

    # Output the ground state energy
    println(energy)
end
\end{lstlisting}

to match the assertion requirements. Then this new script is executed, whose output is further compared to the assertion ground truth. The Scientific verifier implements this procedure for every assertion and generates a comprehensive report. 

\textbf{Error correction. }
Analogous to the programming test, the Scientific verifier first synthesizes the findings from both testing stages into a repair suggestion list, and then executes precise modifications on the script. The updated script then proceeds to the next iteration of checks, repeating this cycle until a complete pass (for both the rubric test and the physical assertion test) is achieved.

\textbf{Convergence test. }
Upon successfully passing the rubric test and the physical assertion test, the Scientific verifier conducts a convergence test to ensure that the computational results of the script are fully convergent. Specifically, the Scientific verifier progressively increases the parameters that govern computational accuracy, such as the number of sweeps in DMRG calculations, and compares the numerical differences between the resulting outputs. Once this difference falls below a predefined criterion, the computation is viewed to be convergent, and the final reproduction results are outputted. 

\end{document}